\documentclass[structabstract]{aa}  
\usepackage{graphicx}
\usepackage{txfonts}
\usepackage{amssymb,amsmath,graphics,multirow,lscape}
\usepackage{natbib}
\bibpunct{(}{)}{;}{a}{}{,}
\usepackage{textcomp}
\usepackage{color}
\usepackage[hang]{subfigure}
\newcommand{\teff}{$T_\mathrm{eff}$}
\newcommand{\logg}{$\log g$}

\newcommand{\water}{H$_2$O}
\newcommand{\invcm}{cm$^{-1}$}
\newcommand{\kms}{km\,s$^{-1}$}
\newcommand{\mic}{$\mu \mathrm m$}

\begin{document}

    \title{Systematic trend of water vapour absorption in red giant atmospheres revealed by high resolution TEXES 12\,\mic\ spectra
       \thanks{M. Richter and T. K. Greathouse were Visiting Astronomers at the Infrared Telescope Facility,
which is operated by the University of Hawaii under Cooperative Agreement
no. NCC 5-538 with the National Aeronautics and Space Administration, Office
of Space Science, Planetary Astronomy Program.}}

\author{N. Ryde\inst{1}  \and J. Lambert\inst{1}   \and M. Farzone\inst{1}  \and M. J. Richter\inst{2} \and  E. Josselin\inst{3}  \and G.M. Harper\inst{4} \and K. Eriksson\inst{5} \and T. K. Greathouse\inst{6}  }

\institute{Department of Astronomy and Theoretical Physics, Lund Observatory, Lund University, Box 43, SE-221 00 Lund, Sweden\\ \email{ryde@astro.lu.se} \and
Department of Physics, University of California at Davis, CA 95616, USA \and 
UPM, Universit\'e Montpellier II, Montpellier, France \and
Astrophysics Research Group, Trinity College Dublin, Dublin 2, Ireland\and
Department of Physics and Astronomy, Uppsala University, Box 516, SE-751 20 Uppsala, Sweden\and
Southwest Research Institute, Division 15, 6220 Culebra Road, San Antonio, TX 78228, USA}

\titlerunning{Water on red giants revealed by TEXES 12 \mic\ spectra}
\authorrunning{N. Ryde et al.}
	    
 \date{Submitted 2014; accepted 2014}

\abstract
   {The structures of the outer atmospheres of red giants are very complex. Recent interpretations of a range of different observations have led to contradictory views of these regions. It is, however, clear that classical model photospheres are inadequate to describe  the nature of the outer atmospheres. The notion of large optically thick  molecular spheres around the stars (MOLspheres) has been invoked in order to explain spectro-interferometric observations, as well as low- and high-resolution spectra. On the other hand high-resolution spectra in the mid-IR do not easily fit into this picture. On the contrary, they rule out any large sphere of water vapour in LTE surrounding red giants. }
   {In order to approach a unified scenario for these outer regions of red giants, more empirical evidence from different diagnostics are needed. Our aim here is to investigate high-resolution, mid-infrared spectra for a range of red giants, spanning spectral types from early-K to mid M. 
We want to study how the pure rotational lines of water vapour change with effective temperature, and whether we can find common properties that can put new constraints on the modelling of these regions, so that we thereby can gain new insights. }
   {We have recorded  mid-infrared spectra at $12.2-12.4\,$\mic\ at high spectral resolution of 10 well-studied bright red giants, with TEXES mounted on the IRTF on Mauna Kea. These stars span effective temperatures from 3450 to 4850 K. }
   {We find that all red giants in our study cooler than 4300 K, spanning a large range of effective temperatures  (down to 3450 K), show water absorption lines stronger than expected  and none are detected in emission, in line with what has been previously  observed for a few stars.  The strengths of the lines vary smoothly with spectral type. We identify several spectral features in the wavelength region that undoubtedly  are formed in the photosphere.  From a study of water-line ratios of the stars, we find that the excitation temperatures, in the line-forming regions, are several hundred Kelvin lower than expected from a classical photospheric model. }
   {All stars in our sample show several photospheric features in their $12\,$\mic\ spectra, which can be modelled with a classical model photosphere. However, in all stars  showing  water-vapour lines (stars cooler than $\sim4300$ K), the water lines are found to be much deeper than expected.  The line ratios of  these pure-rotational lines reveal low excitation temperatures.  This could either be due to an actually lower temperature structure in the outer regions of the photospheres caused by, for example,  extra cooling,  or due to non-LTE level populations, affecting the source function and line opacities, but this needs further investigation. 
We have demonstrated that these diagnostically interesting  water lines are a general feature of red giants across spectral types,  and we argue for a general explanation of their formation rather than  explanations requiring specific properties, such as dust. Since  the water lines are neither weak (filled in by emission) nor appear in emission, as predicted by LTE MOLsphere models in their simplest forms,  the evidence for the existence of such large optically-thick, molecular spheres enshrouding the stars is weakened. It is still a challenge to find a unifying picture of the outer regions of the atmospheres of red giants, but we have presented new empirical evidence that needs to be taken into account and explained in any model of these regions.}

\maketitle

\section{Introduction}

Recent investigations of K and M giants and supergiants have led to a divergent, and sometimes a somewhat contradictory picture of their outer atmospheres. Existing observations range from  interferometric, spectro-interferometric to low- and high-resolution spectra in different bands. Here, we will discuss the picture that has emerged based mostly on near- and mid-infrared data. 

Low-resolution spectra of a range of giants and supergiants \citep[see, for example,][]{tsuji_1997,yam_99,tsuji:2000}  have been interpreted as evidence for a structure, beyond the photosphere, of a molecular condensation of large radial extent, named a MOLsphere \citep{tsuji_1997,tsuji:2000}. \citet{tsuji:08,tsuji:09} concludes, based on high-resolution near-IR spectra, that these structures appear to be a basic feature of all K and M red giants. The structure is supposed to be composed of at least water vapour and CO. It would be a quasi-static molecular layer situated on the order of a stellar radius beyond the classical extent of the photosphere \citep[see for example][]{tsuji:03}. In the case for early K giants like Arcturus, it could be interpreted as an aggregation of molecular clouds connected to the outer photosphere \citep{tsuji:09}. 
Spectro-interferometric K-band observations  have also been interpreted as evidence for molecular layers of water and CO in extended atmospheres of  K and M giants \citep[see for example][]{ohnaka:12,ohnaka:13:atau} and of red supergiants \citep[see for example][]{perrin:04a,perrin:07,Wittkowski:12,ohnaka:13:asco,Arroyo:13}. These spatial extensions can not be modelled with current atmospheric models, indicating that the model atmospheres are too compact \citep{ohnaka:13:atau,Arroyo:13}. 


High-resolution spectra in the mid-infrared of both giants and supergiants have not been able to corroborate this idea.  On the contrary, resolved $12\,$\mic\  spectra  show strong absorption lines of OH and \water, absorption that is actually even larger than that expected from a classical photosphere without a MOLsphere \citep{ryde:water0,ryde:water3,ryde:water2,ryde:water1}. 
Even though it depends on the parameters assumed in the MOLsphere realisation, deep \water\ lines are not expected from an extended, optically-thick MOLsphere, since the water lines will be filled-in with emission from the extended shell  \citep[for an explanation see][]{tsuji:2000,tsuji:03,tsuji:06,ryde:water1}. In short, the reason is that, at wavelengths longward of $\sim5\,$\mic, the cool MOLsphere of {\it large} extent will, in optically thick lines, outshine the hotter stellar continuum from the  geometrically smaller star \citep[see, for the case of $\mu$ Cep,][]{ryde:water1}. This leads to predicted  weak absorption or even emission lines
\citep[see the spectra for a few MOLsphere realisations in][]{ryde:water2}.  The same MOLsphere phenomenon is suggested to be active in normal giants too
\citep{tsuji_2001}. Note that MOLspheres, as molecular clouds closer to the photosphere, will not have the same effect \citep[see][]{tsuji:09}.


 Similarly, but  based on low-resolution infrared spectra, several papers have also noted that synthetic spectra underestimates observed molecular bands. For example, 
\citet{Price:02} demonstrated clearly that synthetic spectra of late 
K-giants observed with the Short Wavelength Spectrometer (SWS) on board the {\it Infrared Space Observatory, ISO} at $R\sim1500$, underpredict the CO bands at $4.5\,$\mic\ and the SiO molecular bands at $8\,$\mic. 
Recently, \citet{Sloan:14} (in press) also showed, based on 
spectra obtained with the Infrared Spectrograph (IRS) on the {\it Spitzer Space Telescope} of 33 K giants at $R\sim100$,  that the observed strengths of the SiO band at $8\,$\mic\  and the OH lines  at $14-17\,$\mic\  are deeper than expected from synthetic spectra, with the interesting finding that the discrepancy for the SiO is smaller than that of
the OH lines, i.e. that it seems to grow with wavelength. They further show that this 
underprediction is an issue for all K giants observed. Likewise, \citet{VanMalderen:04} demonstrated, although more qualitatively, 
the underprediction of, most prominently, the OH lines at $14-17\,$\mic\ also observed with ISO/SWS. There seems to be a consensus that for the outer, tenuous
regions of the atmospheres of red giants where the strong molecular lines form, existing model atmospheres have problems in adequately representing them  \citep{ryde:water0,Price:02,decin:03,farzone:13,Sloan:14}. 
In these regions the heat capacity per volume
is low, which means that even small alterations of the heating or cooling terms in the energy
equation (e.g., due to dynamic processes or uncertainties
and errors in the calculations of radiative cooling) easily can lead
to changes in the temperature structure. 

Thus, based on  a large number of both spectroscopic and interferometric studies of a range of giants and supergiants, it 
is  clear that classical model atmospheres are inadequate to describe the nature of the outer atmospheres of these stars \cite[see, for example,][]{tsuji:03,tsuji:08}. It is, however, unclear whether the physics that these different observations are revealing have a common origin or whether there are several different explanations for different wavelengths, observing method, or different stars in the range of effective temperature and surface gravity that has been investigated. \citet{tsuji:06} concludes that even for the case of the well-studied supergiant Betelgeuse, 
our understanding of the outer atmosphere has not converged. 
The controversy between the different pictures that emerge from the interpretations of interferometric and high-resolution spectroscopic investigations could be a result of the extreme complexity of the outer atmospheres  and could provide a clue to the underlying physical reality \citep{tsuji:06}. 

More data are needed to be added to the bank of empirical evidence in order 
to approach a unified picture. In this paper, we add new insights by investigating the way the 12\,\mic\ \water\ lines 
change with spectral type for a range of K and M giants. This will show whether the larger-than-expected water absorption is  general feature for cool giants across spectral types or due to a specific property, such as the presence of alumina dust, 
as suggested by \citet{tsuji:06,tsuji:08} for Betelgeuse and $\mu$ Cep.

Since the origin of formation of these water lines are not clear, high-resolution spectra are needed to retrieve as much information as possible.   Resolving the stellar spectrum has the advantages that
(1) several resolved \water\ lines of different excitation can be analysed to provide an excitation temperature at the location where they are formed, (2) photospheric features, if any,  can be detected between the water lines, (3) velocity shifts between photospheric features and the water lines can be measured, (4) line broadening can be measured, (5) emission in lines from an extended shell should be present, either as pure emission lines or as emission filling in photospheric absorption lines to different degrees. Detailed line strengths of different resolved lines can also be tested against line-formation models.

With our high-resolution TEXES spectra of a range of K and M giants, we are now in a position to examine these issues.

\section{Observations}


A total of 10 early-K to mid-M giants and sub-giants, ranging from spectral type K0 to M3, were observed in the $12\,$\mic\ region with TEXES, the Texas Echelon Cross Echelle Spectrograph \citep{texes}, mounted on the InfraRed Telescope Facility, IRTF, on Mauna Kea. The observations were performed on 31 Oct., 27 Nov. 2000, 2 \& 4 Feb. 2001, and on 9 Jul 2006. Table \ref{stars} presents the basic data of the observed stars. 

Mid-infrared observations at high spectral resolution have become practical in the past decade due to the development of large-grating spectrographs that take advantage of large-format detector arrays at these wavelengths.
The detector arrays are able to record a weak stellar signal on top of a very bright background. It is, however, still only the very brightest stars that are possible to observe with a decent signal-to-noise ratio within a reasonable observing time.  Indeed, the observed stars are all very bright ($K \lesssim 0$) and lie in the solar neighbourhood, i.e at a distance of less than 100 pc from the Sun.  According to the SIMBAD astronomical database\footnote{http://simbad.u-strasbg.fr/}, there are around 50 stars that meet these criteria. Thus, a fifth of these giants and subgiants in the solar neighbourhood, are represented in 
our data set. Table \ref{stars} presents the bright $K$ and $N$ magnitudes and the final signal-to-noise ratio per pixel of the observations, ranging from $70$ to $290$. The N-band extinction towards the stars are minimal \citep[see for example][]{cardelli}.

The stars were observed with TEXES at high spectral resolution.
The spectral resolution, as determined by Gaussian fits to telluric atmospheric lines, has a FWHM corresponding to $R= 85,000$ or $3.5\,$\kms\  and the slit width was 1.5\arcsec\ and length 8\arcsec. The cross-dispersed spectra consist of approximately 20 orders, each order being larger than the $256^2$ pixel detector array, which results in gaps  between the orders in the final merged spectrum. The spectral coverage ranges from $806.5$ to $821.5\,$\invcm\ (that is approximately  $12.2-12.4\,$\mic). \citet{ryde:water0} describe the determination of the frequency scale, which is subject to systematic offsets of  a half\footnote{$0.5\,$\kms\ corresponds to approximately 0.2\,\AA\ or 0.0014\,\invcm.} to one km\,s$^{-1}$. One pixel represents approximately $0.9\,$\kms. The total integration times of the stars range from $60-600$~\,s. For details about the observing method, see for example, \citet{ryde:water1}. 

\begin{figure}
  \centering
	\includegraphics[width=9cm]{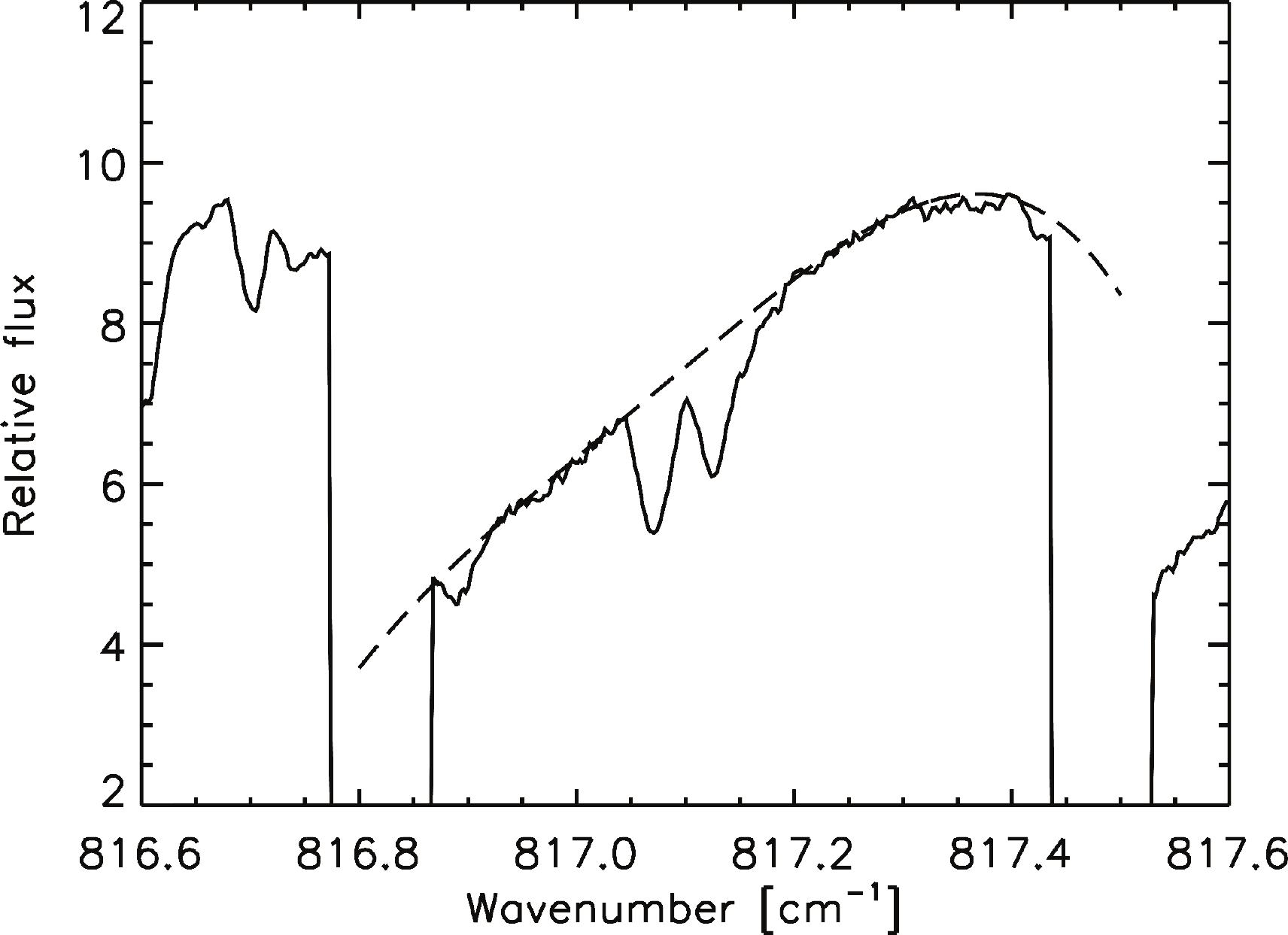}
	\caption{The un-normalized spectrum of $\beta$ Peg around $817.0\,$\invcm. One full order and parts of two neighbouring orders on either side are presented. Three distinct  water lines can be seen in the middle order. The non-linear form of the blaze function is indicated by the dashed line, which is a fifth order Legendre polynomial used to fit the continuum.  } 
	\label{bpeg}
\end{figure}

The observations were reduced using standard TEXES procedures \citep{texes} and the continua are normalised with the \texttt{IRAF}\footnote{IRAF is distributed by the National Optical Astronomy Observatories, which are operated by the Association of Universities for Research in Astronomy, Inc., under cooperative agreement with the National  Science Foundation.} task \texttt{continuum} \citep{IRAF} by a fifth-order Legendre polynomial. Figure \ref{bpeg} shows an example of an un-normalised spectrum, demonstrating the non-linear form of the blaze function. This spectrum of $\beta$ Peg shows two distinct water lines in the middle order.  The normalisation function at the blue edge (at higher wave numbers) is very uncertain due to the turn-over of the blaze function. This should be held in mind when studying features at the blue edges, where extra care should be exercised. The normalised spectrum of $\beta$ Peg is presented later in the top panel of Figure \ref{data1}. 

Subsequently, telluric lines were removed by dividing the normalised spectra with that of a normalised, telluric standard of high signal-to-noise ratio, which we observed in the same setting and reduced in the same way, using custom IDL routines. 
 
Figures \ref{data1} and \ref{data2} present a part of the reduced, normalised spectra (816.5-821.5\,\invcm) in black. The stars are shown in order of increasing effective temperature, \teff. In Figure \ref{data1} these are $\beta$ Peg, $\delta$ Vir,  $\delta$ Oph, $\mu$ UMa, and $\alpha$ Lyn, with effective temperatures between 3448 to 3836\,K. Figure \ref{data2} shows the other stars, namely $\alpha$ Tau, $\alpha$ Hya, $\alpha$ Boo, $\alpha$ Ser, and $\beta$ Gem, with temperature ranging from 3871 to 4858\,K. 

Apart from water lines (we have concentrated our analysis on the ones marked with vertical dashed lines in Figures \ref{data1} and \ref{data2}) several emission lines of Mg, Si, and Ca are detected in the spectra. Figure \ref{em} shows these in detail. Figures \ref{data1} and \ref{data2} also indicates the photospheric OH quartet at $820\,$\invcm\, as well as the rotational line of HF at $819.1\,$\invcm. This latter line is discussed in detail in a paper of its own \citep{joensson:14b}.

\begin{table*}
\caption{K, N magnitudes, parallaxes, distances, and observed signal-to-noise ratios (per spectral pixel) of the program stars\label{stars}}
\centering
\begin{tabular}{ccccccccccc}
\hline  \hline
Name & HR & HD & RA (J2000) & Dec (J2000) &  Spectral & $K$ & $N$  & $\pi^c$ & Distance, $1/\pi$   &{S/N} \\
 & &  & (h:m:s)    & ($\degr$:$\arcmin$:$\arcsec$)  & Type$^a$ &  \multicolumn{2}{c}{magnitudes$^b$} &  [mas] & [pc] & pixel$^{-1}$\\
 \hline
$\beta$ Peg  &     8775 & 217906  & 23:03:46.457 & +28:04:58.03 & M$2.5 \, \mathrm{II}$   & -2.2 &  -2.5  &  $16.64 \pm 0.15$ & $60.1 \pm 0.5$ &   140 \\
$\delta$ Vir &     4910 & 112300  & 12:55:36.209 & +03:23:50.89 & M$3.0 \, \mathrm{III}$  & -1.2 &  -1.5  &  $16.44 \pm 0.22$ & $60.8 \pm 0.8$ &   110 \\
$\delta$ Oph &     6056 & 146051  & 16:14:20.739 & -03:41:39.56 & M$0.5 \, \mathrm{III}$  & -1.2 &  -1.4  &  $19.06 \pm 0.16$ & $52.5 \pm 0.4$ &   90 \\
$\mu$ UMa    &     4069 & 89758   & 10:22:19.740 & +41:29:58.27 & M$0.0 \, \mathrm{III}$  & -0.9 &  -1.0  &  $14.16 \pm 0.54$ & $70.6 \pm 2.7$ &   100 \\
$\alpha$ Lyn &     3705 & 80493   & 09:21:03.301 & +34:23:33.22 & K$7.0 \, \mathrm{III}$  & -0.6 &  -0.8  &  $16.06 \pm 0.17$ & $62.3 \pm 0.7$ &   100 \\
$\alpha$ Tau$^1$ & 1457 & 29139   & 04:35:55.239 & +16:30:33.49 & K$5.0 \, \mathrm{III}$  & -2.8 &  -3.1  &  $48.94 \pm 0.77$ & $20.4 \pm 0.3$ &   90 \\
$\alpha$ Hya$^2$ & 3748 & 81797   & 09:27:35.243 & -08:39:30.96 & K$3.0 \, \mathrm{II}$   & -1.2 &  -1.5  &  $18.09 \pm 0.18$ & $55.3 \pm 0.6$ &   120 \\
$\alpha$ Boo$^3$ & 5340 & 124897  & 14:15:39.672 & +19:10:56.67 & K$1.5 \, \mathrm{III}$  & -3.0 &  -3.2  &  $88.83 \pm 0.54$ & $11.3 \pm 0.1$ &   290 \\
$\alpha$ Ser$^4$ & 5854 & 140573  & 15:44:16.074 & +06:25:32.26 & K$2.0 \, \mathrm{III}$  &  0.1 &  -0.0  &  $44.10 \pm 0.19$ & $22.7 \pm 0.1$ &   70 \\
$\beta$ Gem$^5$  & 2990 & 62509   & 07:45:18.950 & +28:01:34.31 & K$0.0 \, \mathrm{III}$  & -1.1 &  -1.2  &  $96.54 \pm 0.27$ & $10.4 \pm 0.1$ &   80 \\
\hline
\end{tabular}
\tablebib{
(a)~\citet{moz:03}; (b)~Johnson photometric K magnitudes from SIMBAD (http://simbad.u-strasbg.fr/) and N magnitudes from IRAS PSC \citep{iras} with color corrections of 1.41 (see http://lambda.gsfc.nasa.gov/product/iras/colorcorr.cfm); (c) \citet{hipp07}}
\tablefoot{
\tablefoottext{1}{Aldebaran; }
\tablefoottext{2}{Alphard; }\tablefoottext{3}{Arcturus; }\tablefoottext{4}{Unukalhai; }\tablefoottext{5}{Pollux}
}
\end{table*}

\section{Analysis}

In our analysis, we calculate synthetic spectra of the wavelength region 
with the Bsyn code, version 7.09 \citep[for details see][]{ryde:water0}. Bsyn is based on the same routines as the MARCS code \citep{marcs:08}. This latter code is used to construct model atmospheres (in our case in spherical geometry), which   are used when calculating the synthetic spectra. The MARCS code computes hydrostatic model photospheres under the assumption of LTE (Local Thermodynamic Equilibrium), chemical equilibrium, homogeneity and the conservation of the total flux. The fundamental parameters defining a stellar atmosphere model are {\it (i)} the effective temperature, $T_\mathrm{eff}$,  {\it (ii)}  the surface gravity, $\log g$,  {\it (iii)} the metallicity, [Fe/H],  {\it (iv)}  the microturbulence, $\xi_\mathrm{micro}$, and  {\it (v)}  the $\alpha$ element enhancement, [$\alpha$/Fe]. Therefore, we have to determine these parameters for our observed stars, which is done in the following section, Sect. \ref{param}. 

When generating the synthetic spectra with the Bsyn code, the continuous opacities are taken from the model atmosphere calculations. Sect. \ref{line} describes the line data. Subsequently, the spectra were broadened by macroturbulence velocity in order for the synthetic spectrum to fit the observed one. The Eqwi code (version 7.06) calculates equivalent widths for a given model atmosphere and line list (Sect. \ref{line}). Our observed equivalent widths are measured within the \texttt{IRAF} task \texttt{splot}, by measuring the total area in the absorption or emission line.

\subsection{Stellar parameters\label{param}}

\begin{figure}
  \centering
	\includegraphics[width=9.5cm]{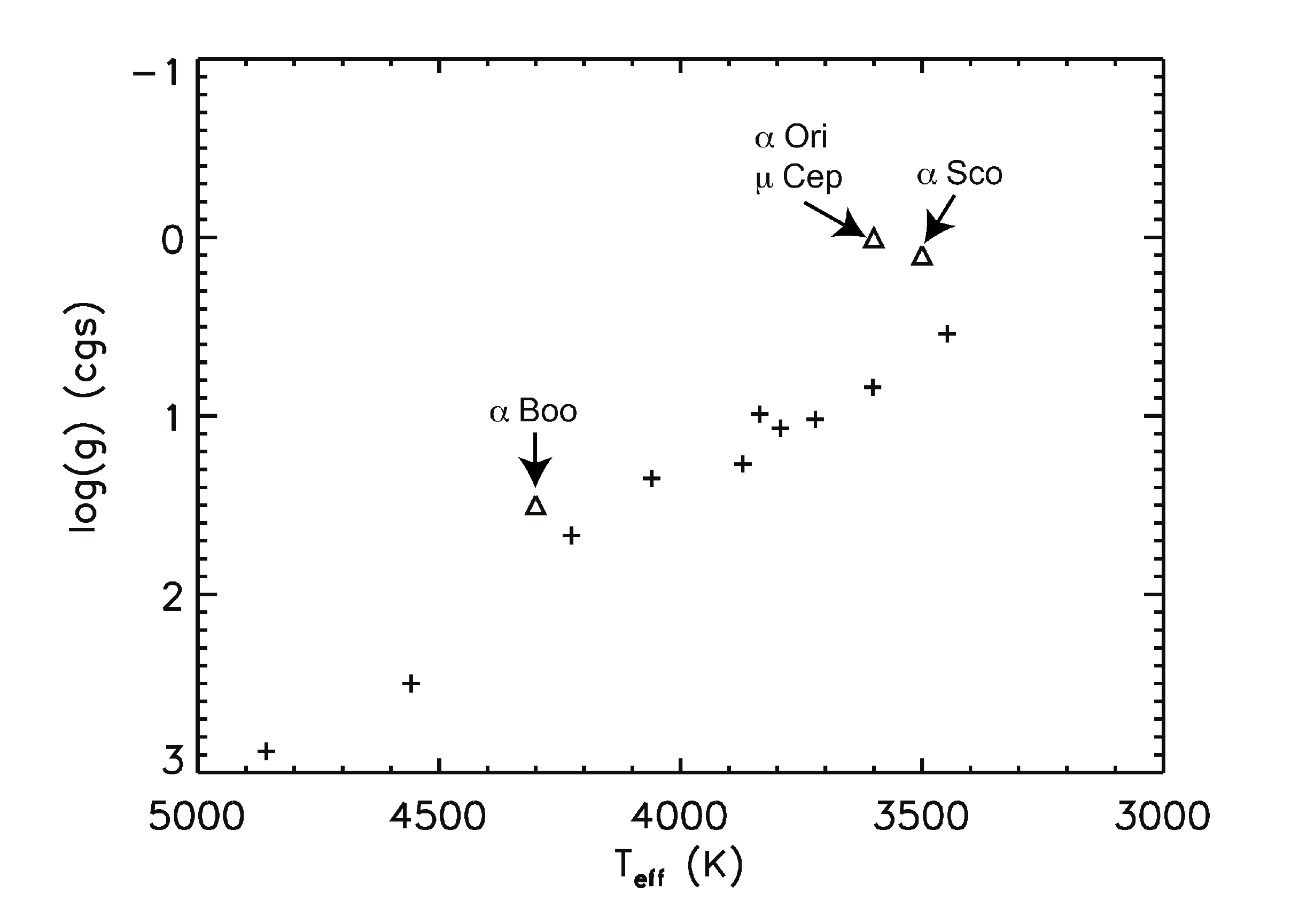}
    \caption{Hertzsrpung-Russell diagram presenting the surface gravities ($\log$g) vs. the effective temperatures ($T_\mathrm{eff}$) for the stars investigated in this paper (shown with plus signs). We have also indicated, with triangles, the supergiants and $alpha$ Boo, that have been investigated at high spectral resolution earlier in the literature \citep{antares,ryde:water0,ryde:water2,ryde:water1}.}
   \label{HR}
\end{figure}

Here we describe our method of determining the stellar parameters, of which the surface gravity is the most uncertain. These are provided in Table \ref{tab:param} and plotted in Figure \ref{HR}. 

\subsubsection{The effective temperature, \teff}
All our stars are bright and nearby. Angular diameters can therefore be acquired with good accuracy. Knowing the bolometric flux, $F_\text{bol}$, and the limb-darkened diameter of a star, $\theta_\text{LD}$, its effective temperature
can be determined directly from
\begin{equation}
	T_\text{eff} = \left(\frac{4F_\text{bol}}{\sigma\theta_\text{LD}^2}\right)^{1/4},
	\label{eq:teff}
\end{equation}

where $\sigma$ is the Stefan Boltzmann constant. \citet{moz:03} measured limb-darkened diameters with the Mark II stellar Interferometer, and provide bolometric fluxes for 85 stars, among others the giants in our sample. They used quadratic limb-darkening laws based 
on ATLAS9 models to derive the diameters. Table \ref{tab:param} gives the effective temperatures and their uncertainties derived by \citet{moz:03}. The uncertainties are of the order of 50 K.

\citet{merand:10} showed that for a K0III star, the uncertainty in 
the diameter due to different models do not exceed 0.5\% 
(and is thus negligible for the determination of $T_\mathrm{eff}$). It may be worse for M stars 
but the uncertainty should not exceed 
$\sim100$\,K. Note that, recently, \citet{arroyo:14}  derived 
an effective temperature of $T_\mathrm{eff}=3909\pm187$\,K for our coolest star $\beta$ Peg
based on {\it VLTI/AMBER} diameters, more than 450 K warmer than the temperature of \citet{moz:03}. 
\citet{dehaes:11} derived $T_\mathrm{eff}=3600\pm300$\,K from modelling the Spectral Energy Distributions, SED, for the same star. For  the warmer stars $\alpha$ Tau and $\alpha$ Boo, they, however, find excellent agreement.
This demonstrates the difficulties for the cooler stars. However, these estimates of the effective temperature of $\beta$ Peg, are higher than the one we adopt, which will only increase the discrepancy between the observed and the modelled \water\ lines in the 12\,\mic\ spectra, which we will discuss later.

\subsubsection{The surface gravity,  \logg, and metalliciy, [Fe/H]}
The surface gravity is given by, 
\begin{equation}
	g = \frac{\text G M}{r^2},
	\label{eq:logg}	
\end{equation}
where G is the gravitational constant, and $r$ the stellar radius. The latter is derived from the measured limb-darkened angular diameter, $\theta_\text{LD}$,  and the distance to the star through its Hipparcos parallax, $\pi$, in the same unit as the angular diameter \citep{hipp07}:	
\begin{equation}
r = 215.1\frac{\theta_\text{LD}}{2\cdot\pi} \  \ R_\odot.
	\label{eq:radii}  
\end{equation}

The stellar mass, $M$, we determine from evolutionary tracks, knowing the stars' luminosity, effective temperature, and metallicity. The stellar metallicities we find from the literature, picking the iron abundances from the references that provide effective temperatures and surface gravities in the proximity of our values, see the references in Table \ref{tab:param}. The literature values can vary (see for example \citet{dehaes:11,cruz:13,joensson:14b}), but will not influence our conclusions.  We use the sets of evolutionary tracks from \citet{Bertelli08, Bertelli09} who use isochrones from  \citet{Girardi00}. We find that the masses vary from $0.9\,\mathrm{M}_\odot$ for $\alpha$ Lyn to $2.7\,\mathrm{M}_\odot$ for $\alpha$ Hya. The uncertainties in the derived luminosities and effective temperatures provide estimates of the uncertainties in the masses, which together with the uncertainties in the radii provide our uncertainties in the surface gravity.  Table \ref{tab:param} summarises the surface gravities and the metallicities, which have typical uncertainties of 0.15 dex, for our stars.

We have also determined the surface gravities for our stars directly from the PARAM 1.1 web interface \citep{daSilva06}. It provides a Bayesian estimate of the stellar parameters $M$, $\log(g)$ and $r$  with the use of the \citet{Girardi00} isochrones, through prior knowledge of the probabilities associated with the placement of a star in the HR-diagram. The input parameters needed are the star's effective temperature, metallicity, $V$-magnitude, and parallax.  Table \ref{tab:param} gives the PARAM gravities in the fourth column.

We can conclude that the surface gravities determined with the two methods agree nicely, within the estimated uncertainties, for all stars except for the two coolest giants. The PARAM gravities are systematically and significantly lower than ours. The only difference in methodology between the two methods is the determination of the stellar luminosity and thereby the stellar radius. PARAM uses the V magnitude and a bolometric correction,  (B.C., from \citet{girardi:08}), which together with the input parameters \teff\ and parallax, $\pi$, provides the radius. We use the angular diameter, parallax, and measured bolometric flux to obtain the luminosity. Although, we are not certain for the cause of the difference in \logg, we note that the main difference is the introduction of the B.C., which gets increasingly larger and more uncertain the cooler the star is. Values from recent literature provide even higher surface gravities, with our \logg\ being closer to these. Although they also find a higher temperature, \citet{Soubiran08}\footnote{They derive the stellar parameters from ELODIE spectra using the TGMET code which minimises a comparison to library spectra.} for instance derive $\log g=1.2$ for $\beta$ Peg, our coolest star, whereas we find $\log g=0.5$ and PARAM provides $\log g=0.2$. In our discussion later about the spectral features in our TEXES spectra, the surface gravity will, however, not play an important role.

\subsubsection{The microturbulence and macroturbulence}

The microturbulence parameter,  $\chi_\mathrm{micro}$, is difficult to determine empirically, but is nevertheless important and  especially so when 
analysing lines that are saturated. The microturbulence will then affect the line strengths for a given abundance, and is therefore a crucial parameter when analysing strong lines. We have, therefore, looked at two different sets of microtubulence determination for giants. One is the values used in the investigations of high-resolution near-infrared spectra of red giants by T. Tsuji \citep[see, for instance, ][]{Tsuji85,tsuji:08} and the second is an empirically found function of $T_\mathrm{eff}$, $\log g$, and [Fe/H] with a scatter of $\sim0.1$~dex, developed within the Gaia-ESO survey \citep{Gilmore12} from high-resolution optical spectra. We present them in different columns in Table \ref{tab:param}.

As we can see from these values, Tsuji's microturbulent velocities are in general 1\,\kms\ larger than the values determined from the empirical Gaia-ESO relation. Since we are not able to determine the microturbulent velocities ourselves, we will take an average of 2.0\,\kms\ as our default value, and later discuss its influence on the spectral lines in the Discussion section, see Sect. \ref{solutions}.
 
To match synthetic spectra with the observed ones, a `macroturbulent' broadening, $\chi_\mathrm{macro}$, is introduced, which takes into account the macroturbulence of the stellar atmosphere and instrumental broadening. Since our water lines are far from  matched by our model spectra, we have convolved our synthesised spectra with reasonable values that match the lines that are modelled well, which are the OH lines and the HF line. We have used a gaussian function, specified by its FWHM which Table~\ref{tab:param} provides. The macroturbulent broadening  changes only the form of a line and not the intrinsic line strength.

\begin{table*}
\caption{Fundamental stellar parameters which are used when generating model atmospheres for our stars \label{tab:param}}
\centering
\begin{tabular}{cccccccccr}
\hline  \hline
 Name  & $T_\textrm{eff}^{(a)}$  & $\log g$$^{(b)}$  & $\log g$ & \textrm{[Fe/H]$^{(b)}$} & $\xi_\mathrm{micro}^{(c)}$ & $\xi_\mathrm{micro}^{(d)}$ & $\xi_\mathrm{micro}$ & $\xi_\mathrm{macro}$ & [$\alpha$/\textrm{Fe}]$^{(e)}$\\
 &  [K] & (cgs) & (cgs)  & & [kms$^{-1}$] & [kms$^{-1}$] & [kms$^{-1}$] & [kms$^{-1}$] &  \\
  & & & PARAM 1.1 & & Gaia-ESO & Tsuji (85,08)  & final & FWHM & \\
 \hline
 $\beta$ Peg & $3448 \pm 42$  & $0.54$ & $0.19 \pm 0.11$ & $-0.11^{(f)}$ &  $1.7 \pm 0.1$ & $2.5 \pm 0.5$  & $2.0 \pm 0.5$   & $5.0$ & $0.00$  \\
 $\delta$ Vir & $3602 \pm 44$ & $0.84$ & $0.67 \pm 0.09$ & $-0.09^{(f)}$ &  $1.6 \pm 0.1$ & $2.5 \pm 0.3$  & $2.0 \pm 0.5$   & $3.0$ & $0.03$  \\
 $\delta$ Oph & $3721 \pm 47$ & $1.02 $ & $0.89 \pm 0.10$ & $-0.03^{(g)}$ &  $1.6 \pm 0.1$ & $2.5 \pm 0.5$  & $2.0 \pm 0.5$   & $5.0$ & $0.00$  \\
 $\mu$ UMa    & $3793 \pm 47$ & $1.07$ & $0.98 \pm 0.08$ & $0.00^{(h)}$  &  $1.5 \pm 0.1$ & -              & $2.0 \pm 0.5$   & $6.0$     & $0.00$  \\
 $\alpha$ Lyn & $3836 \pm 47$ & $0.98$ & $1.02 \pm 0.06$ & $-0.26^{(i)}$ &  $1.8 \pm 0.1$ & -              & $2.0 \pm 0.5$   & $5.0 $     & $0.08$  \\
 $\alpha$ Tau & $3871 \pm 48$ & $1.27 $ & $1.19 \pm 0.10$ & $0.00^{(f)}$  &  $1.5 \pm 0.1$ & $3.0 \pm 1.0 $ & $2.0 \pm 0.5$   & $5.0 $ & $0.00$  \\
 $\alpha$ Hya& $4060 \pm 50$  & $1.35 $ & $1.20 \pm 0.08$ & $-0.12^{(i)}$ &  $1.5 \pm 0.1$ & -              & $2.0 \pm 0.5$   & $5.0 $    & $0.00$ \\
 $\alpha$ Boo & $4226 \pm 53$ & $1.67 $ & $1.61 \pm 0.05$ & $-0.60^{(j)}$ &  $1.4 \pm 0.1$ & $2.2 \pm 0.2$  & $2.0 \pm 0.5$   & $6.0 $ & $0.18$  \\
 $\alpha$ Ser & $4558 \pm 56$ & $2.50$ & $2.40 \pm 0.10$ & $0.03^{(i)}$  &  $1.2 \pm 0.1$ & -              & $2.0 \pm 0.5$   & $5.0$     & $0.00$  \\
 $\beta$ Gem  & $4858 \pm 60$ & $2.88 $ & $2.79 \pm 0.10$ & $-0.07^{(i)}$ &  $1.0 \pm 0.1$ & -              & $2.0 \pm 0.5$   & $5.0$    & $0.00$  \\
 \hline
\end{tabular}
\tablebib{(a)~\citet{moz:03}; 
(b) typical uncertainty of $\pm 0.15$; (c) Maria Bergemann, private comm.;
(d) based on \citet{Tsuji85,tsuji:08}
(e) see text; (f) \citet{smith:85};  (g) \citet{koleva:12}; (h) \citet{mallik:98}; (i) \citet{mcwilliam:90};
(j) \citet{leep:87}.
}
\end{table*}

\subsubsection{[$\alpha$/Fe] and CNO abundances}

A few of the $\alpha$ elements\footnote{The $\alpha$ elements are 
those from O to Ti with even atomic numbers, i.e. O, Ne, Mg, Si, S, Ar, Ca, and Ti.} are important electron donors in the continuum- and line-forming regions of the stellar atmosphere. Especially, Mg alone contributes more than half of the electrons, but also Si is important at least in the continuum forming regions. The continuous opacity in the mid-infrared is due to  H$^-_\mathrm{ff}$,\footnote{The subscript in H$^-_\mathrm{ff}$ stands for `free-free'.} an opacity which is proportional to the electron density \citep[see, for example, Eq. 2.76 from][]{rutten:03}. The  strength of a weak line is proportional to the line opacity and inversely proportional to the continuous opacity, thus directly influenced by the electron density.

In our atmospheric models and spectral synthesis we have prescribed an $\alpha$ element  over-abundance which is $+0.4$ for metallicities below [Fe/H]$=-1$ and linearly decreasing down to a solar ratio at solar metallicity.

Since we are modelling red giants and the first dredge-up that these stars have experienced will lead to an increased C abundance and an accompanied decrease of N, such that C+N is constant, we have used CN-processed models \citep[see, for example,][]{marcs:08}. However, as these authors point out, the effects on the model structure is moderate.


\subsection{Line data\label{line}}

\begin{table*}

\caption{Molecular data of the most prominent H$_2$O lines in the 808$-$822 cm$^{-1}$ region ($\lambda_\mathrm{air}=12.162-12.373$\,\mic), partly from \citet{ryde:water0} data\label{tab:WaterMolData} }
\centering
\begin{tabular}{c||ccccc}
 \hline \hline
$\nu_\textrm{lab}$   & $E''_\textrm{exc}$ & $\log gf$ & $J'(K_{a}', K_{c}')$ & $J''(K_{a}'', K_{c}'')$ & $v_1v_2v_3$ \\
\,[cm$^{-1}$] & [eV] & & & &  \\
 \hline
808.632 & 1.140 & -1.40 & 23(11,12) & 22(10,13) & (010) \\
814.675 & 1.385 & -1.18 & 20(16,5) & 19(15,5) & (020) \\
814.675\tablefootmark{a} & 1.385 & -1.66 & 20(16,5) & 19(15,5) & (020) \\
815.301 & 0.498 & -2.51 & 18(7,12) & 17(4,13) & (000) \\
815.897\tablefootmark{a} & 1.396 & -1.00 & 21(21,0) & 20(20,1) & (010) \\
815.900 & 1.396 & -1.48 & 21(21,0) & 20(20,1) & (010) \\
816.450 & 0.398 & -3.21 & 17(5,13) & 16(2,14) & (000) \\
816.687 & 1.014 & -1.35 & 24(12,13) & 23(11,12) & (000) \\
817.157 & 1.206 & -1.18 & 21(16,5) & 20(15,6) & (010) \\
817.209 & 1.014 & -1.83 & 24(12,12) & 23(11,13) & (000) \\
818.424\tablefootmark{b} & 1.029 & -1.66 & 22(16,6) & 21(15,7) & (000) \\
818.425\tablefootmark{b} & 1.029 & -1.19 & 22(16,7) & 21(15,6) & (000) \\
819.046 & 1.319 & -1.30 & 21(13,8) & 20(12,9) & (020) \\ 
819.932 & 1.050 & -1.42 & 25(11,14) & 24(10,15) & (000) \\
820.190 & 1.360 & -1.04 & 21(20,1) & 20(19,2) & (010) \\
820.190\tablefootmark{a} & 1.360 & -1.52 & 21(20,1) & 20(19,2) & (010) \\
820.583 & 1.245 & -1.15 & 21(17,5) & 20(16,5) & (010) \\
820.583\tablefootmark{a} & 1.245 & -1.62 & 21(17,5) & 20(16,5) & (010) \\
 \hline
\end{tabular}
\tablefoot{\\
\tablefoottext{a}{Uncertain assignment of the quantum numbers for the states of the transition}\\
\tablefoottext{b}{Assigned from \citet{tsuji:2000}}
}
\end{table*}

Spectral lines detected in the recorded wavelength range include  \water, OH, HF, and metallic emission lines 
(for the latter, see Sect. \ref{emlines}). When synthesising the modelled spectra, we have included an atomic line list and the most common molecules
that might have transitions in the N band.\footnote{We have thus included an atomic line list, lists for \water\, $^{16}$OH, HF, $^{12}$CH, $^{13}$CH, $^{12,13}$C$^{12,13}$C, $^{12,13}$C$^{14,15}$N, $^{12,13}$C$^{16,17,18}$O, $^{28,29,30}$SiO, $^{13}$HCN, and $^{56}$FeH.} For our temperature range, however, only \water\, OH, HF, and atoms show features detectable at the signal-to-noise level of the observations. Neither are atomic absorption lines detected nor are there any expected from a spectral synthesis of the wavelength region. The only atomic lines that are detected are
emission lines due to a very specific non-LTE process, see Sect. \ref{emlines}.  

The line lists for our calculation of the photospheric synthetic spectra  are taken from the compilation of \citet{gold} for OH and from \citet{joensson:14b} for HF.
The \water\ line list is based on the compilation of \citet{Barber06}, but for a dozen of the strongest lines,  \citet{ryde:water0} identified the transitions from laboratory measurements by \citet{poly_1, poly_2,poly_3}. Uncertainties in the line positions of these measurements are less than 0.002\,\invcm\ \citep[for details see][]{ryde:water0,ryde:water2,ryde:water1}. Table \ref{tab:WaterMolData} gives the line data for the most prominent \water\ lines. All molecular lines are synthesised with the Bsyn code with corresponding partition functions that 
are consistent with the line lists.


\section{Results}

\begin{figure*}
  \centering
	\includegraphics[width=17cm]{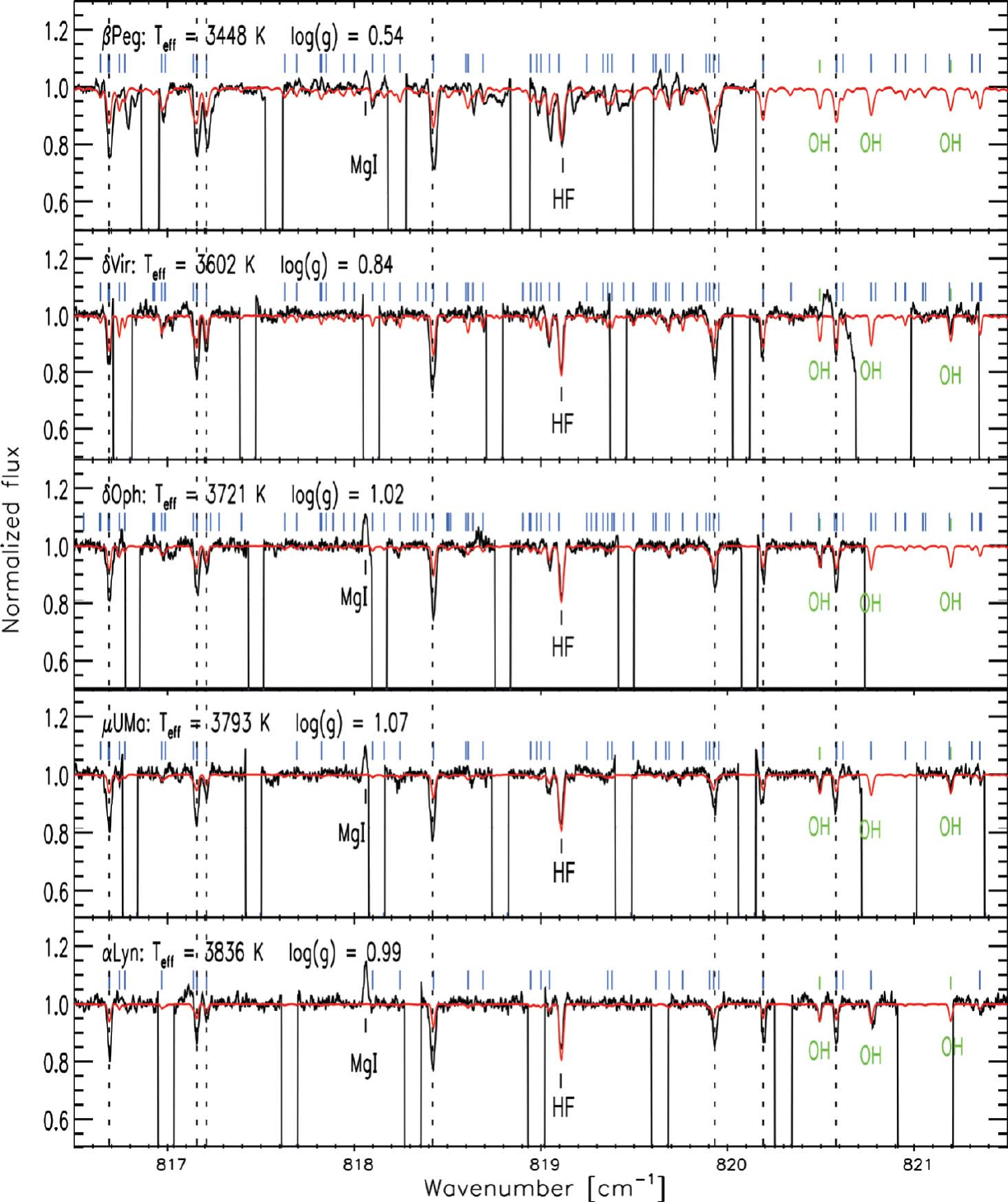}
	\caption{The observations of the five coolest stars. The observed spectra are shown with black lines and the synthetic spectra with red lines. 
The blue ticks above the spectra indicate spectral features of \water\ whereas green marks OH($\nu=3-3$) lines. Dashed lines mark the spectral features of interest which are ultimately used in the calculation of the formation region temperatures. Also, one emission feature in the form of neutral Mg is visible at 818.1\,\invcm, and the HF line is also marked. The lack
of observational data between the orders of the spectrometer, is indicated by the vertical solid black lines.} 
	\label{data1}
\end{figure*}

\begin{figure*}
  \centering
	\includegraphics[width=17cm]{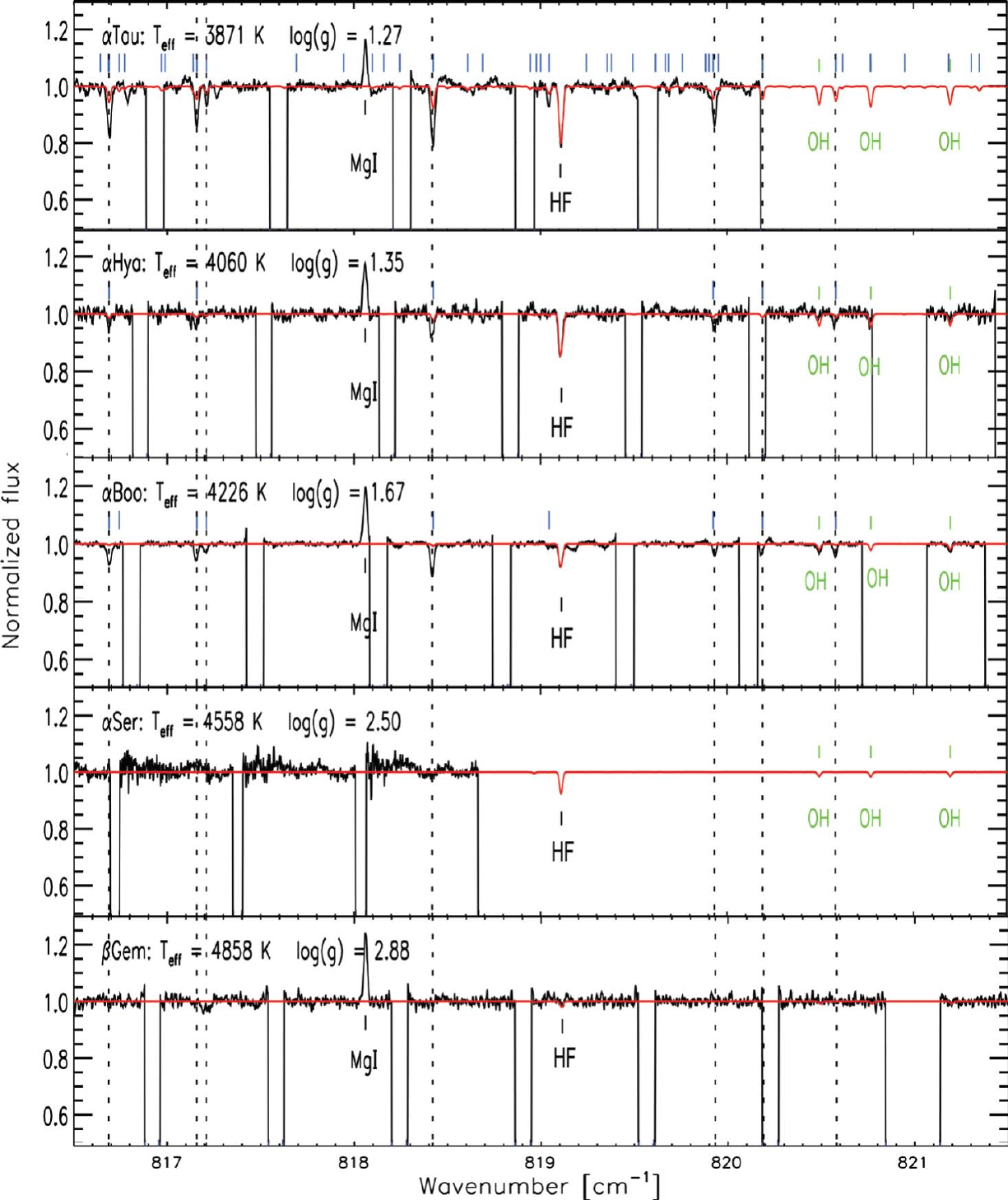}
	\caption{The observations of the five warmest stars. The observed spectra are shown with black lines and the synthetic spectra with red lines. See also figure caption of Figure \ref{data1}.} 
	\label{data2}
\end{figure*}

In Figures \ref{data1} and \ref{data2}, parts of the TEXES spectra of our red giants are shown. The observations are presented in black and the synthetic spectra in red. Water lines that affect the synthetic spectra are marked with blue tick marks and the most prominent water lines, that we will be using in the analysis, are marked with vertical, dashed lines. Absorption lines of OH are shown with green tick marks, and the HF line and the emission line due to Mg are marked. 
From the spectra, no molecular emission lines are detected. Furthermore, the strengths of the \water\ lines vary smoothly with effective temperature and, as expected, the coolest stars have the strongest and most \water\ lines present.  The wavelengths of the strongest water lines are accurate enough for our spectra, whereas several weaker lines do not match up as well. This is expected based on the accuracy of the \water\ line list, see Section \ref{line}. In the following sections, we will be more interested in the lines that appear over a range of stellar temperatures, that is the strongest lines only, those marked with  vertical, dashed lines. 

All lines in the region are modelled and synthesised, except for the metallic emission lines that require a specific non-LTE process (see Section \ref{emlines}), which are not modelled for here \citep[for a model of these lines, see][]{sundqvist:08}. 
From Figures \ref{data1} and \ref{data2} and the other parts of our recorded TEXES spectra, we immediately see that  the synthetic spectra fail to match the observations of the OH($\nu=0-0$) quartets, failing by a bit, and the  \water\ lines,  failing by a large amount. 
The OH($\nu=2-2$), OH($\nu=3-3$), and the HF lines are, however, modelled well with a photospheric model with the stellar parameters derived in section \ref{param}.

{\it This result for all our stars is identical to the results and conclusions for Arcturus by \citet{ryde:water0}.} Now, we see the same trend for all the giants in our investigation, stars with spectral types ranging from K to M.

Table \ref{tab:eqw1} presents our measured equivalent widths of the water lines, $W_\mathrm{obs}$ (in m\AA), and their strengths in the form of $\log (W_\mathrm{obs}/\lambda)$.
We see that the lines are normally weaker than $\log (W_\mathrm{obs}/\lambda)=-5.0$ and that they vary smoothly with the effective temperatures of the stars. 
In the table we also present the modelled equivalent widths, $W_\mathrm{mod}$ (also in m\AA), for the water lines. Also, here we can conclude that the synthetic lines are too weak for all stars and all lines. For the coolest star, $\beta$ Peg, the synthetic lines are too weak by a factor of $\sim$two, and for the warmer stars, a factor five to 50.

\onltab{
\begin{table*}
\caption{The equivalent widths of water lines in m\AA\ as observed and the equivalent widths from synthetic spectra for the stars calculated for standard {\sc marcs} models. A measure of the lines' observational line strength calculated as $\log W_{obs}/{\lambda}$ is also given. $W_\text{obs}$ is the observed equivalent width and $\lambda$ the wavelength.\label{tab:eqw1}}
\centering
\begin{tabular}{ccc||c|c|c||c|c|c||c|c|c}
 \hline \hline
\multicolumn{1}{c}{} & \multicolumn{1}{c}{} & \multicolumn{1}{c}{} & \multicolumn{3}{c}{$\mathbf{808.632 \ \textbf{cm}^{-1}}$} & \multicolumn{3}{c}{$\mathbf{814.675 \ \textbf{cm}^{-1}}$} & \multicolumn{3}{c}{$\mathbf{815.300 \ \textbf{cm}^{-1}}$} \\
\multicolumn{1}{c}{\textbf{HR}} & \multicolumn{1}{c}{\textbf{Name}} & \multicolumn{1}{c}{$T\mathbf{_{eff}}$} & \multicolumn{1}{c}{$W_\mathrm{mod}$} & \multicolumn{1}{c}{$W_\mathrm{obs}$} & \multicolumn{1}{c}{$\log ({W_\mathrm{obs}}/{\lambda})$} & \multicolumn{1}{c}{$W_\mathrm{mod}$} & \multicolumn{1}{c}{$W_\mathrm{obs}$} & \multicolumn{1}{c}{$\log ({W_\mathrm{obs}}/{\lambda})$} & \multicolumn{1}{c}{$W_\mathrm{mod}$} &  \multicolumn{1}{c}{$W_\mathrm{obs}$} & \multicolumn{1}{c}{$\log ({W_\mathrm{obs}/\lambda})$} \\
\multicolumn{1}{c}{} & \multicolumn{1}{c}{} & \multicolumn{1}{c}{[K]} & \multicolumn{1}{c}{[m\AA ]} & \multicolumn{1}{c}{[m\AA ]} &\multicolumn{1}{c}{} &  \multicolumn{1}{c}{[m\AA ]} & \multicolumn{1}{c}{[m\AA ]} &\multicolumn{1}{c}{} & \multicolumn{1}{c}{[m\AA ]} & \multicolumn{1}{c}{[m\AA ]} & \multicolumn{1}{c}{}\\
\hline
HR8775 & $\beta$ Peg & $3448 \pm 42$ & 506 & -- & -- & 598 & 648 & -5.28 & 511 & -- & -- \\
HR4910 & $\delta$ Vir & $3602 \pm 44$ & 324 & 612 & -5.31 & 359 & 301 & -5.61 & 301 & 873 & -5.15 \\
HR6056 & $\delta$ Oph & $3721 \pm 47$ & 211 & 444 & -5.45 & 216 & 301 & -5.61 & 135 & 858 & -5.16 \\
HR4069 & $\mu$ UMa & $3793 \pm 47$ & 143 & 459 & -5.44 & 136 & 331 & -5.57 & 135 & 752 & -5.21 \\
HR3705 & $\alpha$ Lyn & $3836 \pm 47$ & 114 & 321 & -5.59 & 105 & -- & -- & 94.8 & 617 & -5.30 \\
HR1457 & $\alpha$ Tau & $3871 \pm 48$ & 106 & -- & -- & 101 & -- & -- & 91.8 & 662 & -5.27 \\
HR3748 & $\alpha$ Hya & $4060 \pm 60$ & 23.9 & -- & -- & 22.2 & -- & -- & 18.1 & 141 & -5.94 \\
HR5340 & $\alpha$ Boo & $4226 \pm 53$ & 6.40 & 113 & -6.04 & 4.52 & 150 & -5.91 & 4.36 & 196 & -5.79 \\
HR5854 & $\alpha$ Ser & $4558 \pm 56$ & -- & -- & -- & -- & $0.00$ & $-\infty$ & $0.00$ & $0.00$ & 0.00 \\
HR2990 & $\beta$ Gem & $4858 \pm 60$ & -- & $0.00$ & $-\infty$ & -- & $0.00$ & $-\infty$ & $0.00$ & $0.00$ & 0.00 \\
 \hline
\multicolumn{1}{c}{} & \multicolumn{1}{c}{} & \multicolumn{1}{c}{} & \multicolumn{3}{c}{$\mathbf{815.900 \ \textbf{cm}^{-1}}$} & \multicolumn{3}{c}{$\mathbf{816.450 \ \textbf{cm}^{-1}}$} & \multicolumn{3}{c}{$\mathbf{816.687 \ \textbf{cm}^{-1}}$} \\
 \hline
HR8775 & $\beta$ Peg & $3448 \pm 42$ & 713 & 826 & -5.17 & 302 & 765 & -5.20 & 703 & 1290 & -4.98 \\
HR4910 & $\delta$ Vir & $3602 \pm 44$ & 446 & 555 & -5.34 & 150 & 495 & -5.39 & 455 & 1012 & -5.08 \\
HR6056 & $\delta$ Oph & $3721 \pm 47$ & 285 & 466 & -5.42 & 76.0 & 420 & -5.46 & 312 & 884 & -5.14 \\
HR4069 & $\mu$ UMa & $3793 \pm 47$ & 186 & 421 & -5.46 & 38.7 & 285 & -5.63 & 221 & 930 & -5.12 \\
HR3705 & $\alpha$ Lyn & $3836 \pm 47$ & 147 & 481 & -5.40 & 27.7 & 255 & -5.68 & 186 & 705 & -5.24 \\
HR1457 & $\alpha$ Tau & $3871 \pm 48$ & 140 & 421 & -5.46 & 26.4 & 255 & -5.68 & 173 & 750 & -5.21 \\
HR3748 & $\alpha$ Hya & $4060 \pm 60$ & 32.2 & 165 & -5.87 & 5.10 & -- & -- & 43.5 & 225 & -5.74 \\
HR5340 & $\alpha$ Boo & $4226 \pm 53$ & 6.62 & 146 & -5.92 & -- & 67.1 & -6.25 & 11.0 & 315 & -5.59 \\
HR5854 & $\alpha$ Ser & $4558 \pm 56$ & -- & $0.00$ & $-\infty$ & -- & $0.00$ & $-\infty$ & -- & $0.00$ & $-\infty$  \\
HR2990 & $\beta$ Gem & $4858 \pm 60$ & -- & $0.00$ & $-\infty$ & -- & $0.00$ & $-\infty$ & -- & $0.00$ & $-\infty$ \\
 \hline 
\multicolumn{1}{c}{} & \multicolumn{1}{c}{} & \multicolumn{1}{c}{} & \multicolumn{3}{c}{$\mathbf{817.157 \ \textbf{cm}^{-1}}$} & \multicolumn{3}{c}{$\mathbf{817.209 \ \textbf{cm}^{-1}}$} & \multicolumn{3}{c}{$\mathbf{818.425 \ \textbf{cm}^{-1}}$} \\
 \hline
HR8775 & $\beta$ Peg & $3448 \pm 42$ & 612 & 1228 & -5.00 & 610 & 1138 & -5.03 & 1025 & 1612 & -4.88 \\
HR4910 & $\delta$ Vir & $3602 \pm 44$ & 395 & 1063 & -5.06 & 356 & 464 & -5.42 & 694 & 1269 & -4.98 \\
HR6056 & $\delta$ Oph & $3721 \pm 47$ & 263 & 764 & -5.20 & 207 & 374 & -5.51 & 492 & 1120 & -5.04 \\
HR4069 & $\mu$ UMa & $3793 \pm 47$ & 184 & 884 & -5.14 & 126 & 389 & -5.50 & 365 & 985 & -5.09 \\
HR3705 & $\alpha$ Lyn & $3836 \pm 47$ & 150 & 659 & -5.27 & 95.8 & 299 & -5.61 & 311 & 1015 & -5.08 \\
HR1457 & $\alpha$ Tau & $3871 \pm 48$ & 141 & 524 & -5.37 & 91.2 & 225 & -5.74 & 289 & 836 & -5.16 \\
HR3748 & $\alpha$ Hya & $4060 \pm 60$ & 30.9 & -- & -- & 19.4 & -- & -- & 75.6 & 343 & -5.55 \\
HR5340 & $\alpha$ Boo & $4226 \pm 53$ & 8.39 & 240 & -5.71 & 3.64 & 135 & -5.96 & 20.6 & $0.00$ & -5.51 \\
HR5854 & $\alpha$ Ser & $4558 \pm 56$ & -- & $0.00$ & $-\infty$ & -- & $0.00$ & $-\infty$ & 2.65 & $0.00$ & $-\infty$ \\
HR2990 & $\beta$ Gem & $4858 \pm 60$ & -- & $0.00$ & $-\infty$ & -- & $0.00$ & $-\infty$ & -- & $0.00$ & $-\infty$ \\
 \hline
\multicolumn{1}{c}{} & \multicolumn{1}{c}{} & \multicolumn{1}{c}{} & \multicolumn{3}{c}{$\mathbf{819.046 \ \textbf{cm}^{-1}}$} & \multicolumn{3}{c}{$\mathbf{819.932 \ \textbf{cm}^{-1}}$} & \multicolumn{3}{c}{$\mathbf{820.583 \ \textbf{cm}^{-1}}$} \\
 \hline
HR8775 & $\beta$ Peg & $3448 \pm 42$ & 420 & 655 & -5.27 & 505 & 1562 & -4.89 & 833 & -- & -- \\
HR4910 & $\delta$ Vir & $3602 \pm 44$ & 256 & 537 & -5.36 & 338 & 1071 & -5.06 & 527 & 787 & -5.19 \\
HR6056 & $\delta$ Oph & $3721 \pm 47$ & 158 & 313 & -5.59 & 236 & 684 & -5.25 & 342 & 683 & -5.25 \\
HR4069 & $\mu$ UMa & $3793 \pm 47$ & 102 & 343 & -5.55 & 170 & 684 & -5.25 & 229 & 564 & -5.33 \\
HR3705 & $\alpha$ Lyn & $3836 \pm 47$ & 79.5 & -- & -- & 141 & 640 & -5.28 & 184 & 549 & -5.35 \\
HR1457 & $\alpha$ Tau & $3871 \pm 48$ & 75.6 & 209 & -5.77 & 131 & 580 & -5.32 & 173 & -- & -- \\
HR3748 & $\alpha$ Hya & $4060 \pm 60$ & 16.2 & -- & -- & 31.5 & -- & -- & 39.4 & -- & -- \\
HR5340 & $\alpha$ Boo & $4226 \pm 53$ & 4.37 & 118 & -6.02 & 8.43 & 149 & -5.91 & 10.7 & 178 & -5.83 \\
HR5854 & $\alpha$ Ser & $4558 \pm 56$ & -- & -- & -- & -- & -- & -- & -- & -- & -- \\
HR2990 & $\beta$ Gem & $4858 \pm 60$ & -- & $0.00$ & $-\infty$ & -- & $0.00$ & $-\infty$ & -- & $0.00$ & $-\infty$ \\
 \hline
\end{tabular}
\end{table*}
}

\subsection{The formation of the continuum at $12\,$\mic}\label{cont}

None of the red giants we have observed are expected to have large continuum emission due to dust in the 12\,\mic\ region.  This is verified from low spectral-resolution SWS01 spectra observed with the {\it Infrared Space Observatory, ISO} \citep{kessler}. We therefore assume that the continua recorded from the red giants originate in  their photospheres.  For an Arcturus model atmosphere, the near- and mid-IR continua, which are dominated by the H$^-_\mathrm{ff}$ opacity, is formed at an optical depth of $\log\tau_{500}\sim=-0.5$. This can be compared with the continuum at $1.6\,\mu$m, which is formed at $\log\tau_{500}\sim=+0.6$, and is the deepest point in the atmosphere the optical/near-IR continuum is formed.  Even a MOLsphere  would not contribute much continuous opacity \citep[see, for example, the realisations of a MOLsphere in][]{ryde:water1}. {|bf  The MOLsphere continuum, due to H$_\mathrm{ff}$ and  H$^-_\mathrm{ff}$, is therefore expected to be transparent. Other stars, such as the supergiants $\alpha$ Ori and $\mu$ Cep, show large dust emission in the wavelength region of 9-13\,\mic. In these cases this has to properly be taken into account when synthesising this spectral region \citep[see][]{ryde:water2,ryde:water1}. 

\subsection{Emission lines of Mg, Al, Si, and Ca}\label{emlines}

In the parts of our TEXES spectra in Figure \ref{data1} and \ref{data2}, emission lines due to Mg~{\sc i} at 818.06\,\invcm\   are clearly detected in all the stars, apart from those where the line falls into a gap between orders.

\citet{mcmg} explained in detail the formation process of the emission lines of Mg~{\sc i} in the solar spectrum. The  lines are formed in the photosphere due to a non-LTE process, in which  a flow cycle between  states of Mg~{\sc i} and Mg~{\sc ii}, refilling neutral magnesium again via its Rydberg states, is established. This results in an underpopulation of the levels from which the lines originate. A small difference in the departure coefficients between the upper and lower levels is the direct cause of the emission.  

\citet{ryde:04_letter} were the first to report these emission lines in a star apart from the Sun. They  showed that the same non-LTE mechanism is at play in Procyon (F5IV-V), a star with an effective temperature of \teff = 6512 K and a surface gravity of \logg =3.9 (cgs). Later, \citet{sundqvist:08} detected even stronger emission lines in three K giants. They succeeded in modelling these lines by using the same non-LTE mechanism  also for these giants. An expanded model atom of Mg and inclusion of neutral hydrogen collisions was, however, needed in order to reproduce the emission strengths in these giants.  \citet{sundqvist:08} also showed that the emission lines are formed in atmospheric layers below the temperature minimum. They are therefore clearly photospheric features. 

Table \ref{tab:em} gives possible metallic emission lines in our wavelength region, in addition to the Mg~{\sc i} emission lines. \citet{mcmg} and \citet{baumuller} explained and modelled the emission lines due to Si~{\sc i} and Al~{\sc i}  in the solar spectrum, which are weaker than the Mg emission, with a similar process as for magnesium. 
Stellar emission lines at 12\,\mic\ due to Si~{\sc i} and Al~{\sc i} were first identified in $\alpha$ Boo by \citet{texes} and analysed by \citet{sundqvist:08} in the three K giants they investigated.  Here, we detect emission lines due to Si~{\sc i}, Al~{\sc i}, and  Ca~{\sc i}\ lines for all our giants. Figure \ref{em}
presents these  and Table \ref{tab:em:eqw} gives their equivalent widths (when measurable). 
The metallic emission lines at 12\,\mic\ thus seems to be a common feature in a variety of stars, and can be explained by the non-LTE flow cycle in the stellar photospheres. For red giants the lines are formed at an optical depth of $\log \tau_{500}\sim-1.7$, i.e. quite deep in. Is it then clear that we are recording the photospheric spectrum of the red giants.

\begin{table}
\caption{Photospheric emission lines with line identifications from \citet{chang83,chang84,brault83,ryde:04_mg,sundqvist:08}}\label{tab:em}
\centering
\begin{tabular}{lcc}
\hline  \hline

 Element  & wavenumber  & wavelength \\
  & & in air \\
  & [cm$^{-1}$] & [$\mu$m]\\
  
 \hline
Si\,\sc{i}         & 810.344 & 12.3371     \\
Si\,\sc{i}         & 810.360 & 12.3368    \\
Si\,\sc{i}         & 810.591 & 12.3333    \\
Al\,\sc{i}         & 810.704 & 12.3316    \\
Mg\,\sc{i}$^{(a)}$ & 811.578 & 12.3183    \\
Si\,\sc{i}         & 811.709 & 12.3163    \\
Si\,\sc{i}         & 813.380 & 12.2910   \\
Si\,\sc{i}$^{(b)}$ & 814.273 & 12.2775    \\
Ca\,\sc{i}         & 814.969 & 12.2671    \\
Al\,\sc{i}$^{(b)}$ & 815.375 & 12.2610    \\
Si\,\sc{i}$^{(b)}$ & 815.979 & 12.2519    \\
Mg\,\sc{i}         & 818.058 & 12.2207    \\
 \hline
\end{tabular}
\tablefoot{
\tablefoottext{a}{Blended with an absorption line from H$_2$O}\\
\tablefoottext{b}{Blended with an absorption line from OH}
}
\end{table}

\begin{table}
\caption{Equivalent widths in m\AA\ of the photospheric emission lines identified with their wavenumber in cm$^{-1}$. }\label{tab:em:eqw}
\centering
\begin{tabular}{lccccc}
\hline  \hline
 Star  & Si\,\sc{i}  &  Si\,\sc{i} & Al\,\sc{i}  & Ca\,\sc{i} &  Mg\,\sc{i} \\
  & 810.35 & 810.59 & 810.70 & 814.97 & 818.06 \\
 \hline
 $\beta$ Peg    &  &  &  &  & 237.9   \\
 $\delta$ Vir   &  &  &  & 223.8 &  \\
 $\delta$ Oph   &  &  &  & 169.9 & 416.2  \\
 $\mu$ UMa      &  &  &  & 104.2 & 272.1 \\
 $\alpha$ Lyn   &  &  & 235.9 &  & 499.6  \\
 $\alpha$ Tau   &  &  &  &  & 548.9  \\
 $\alpha$ Hya   & 251.5 & 111.1 & 214.6 & 104.2 & 718.1  \\
 $\alpha$ Boo   & 232.2 & 103.5 & 219.1 & 112.2 &  737.0 \\
 $\alpha$ Ser   &  &  &  & 139.6 &  \\
 $\beta$ Gem    &  &  &  &  & 888.8    \\

 \hline
\end{tabular}

\end{table}

\begin{figure*}
  \centering
	\includegraphics[width=18cm]{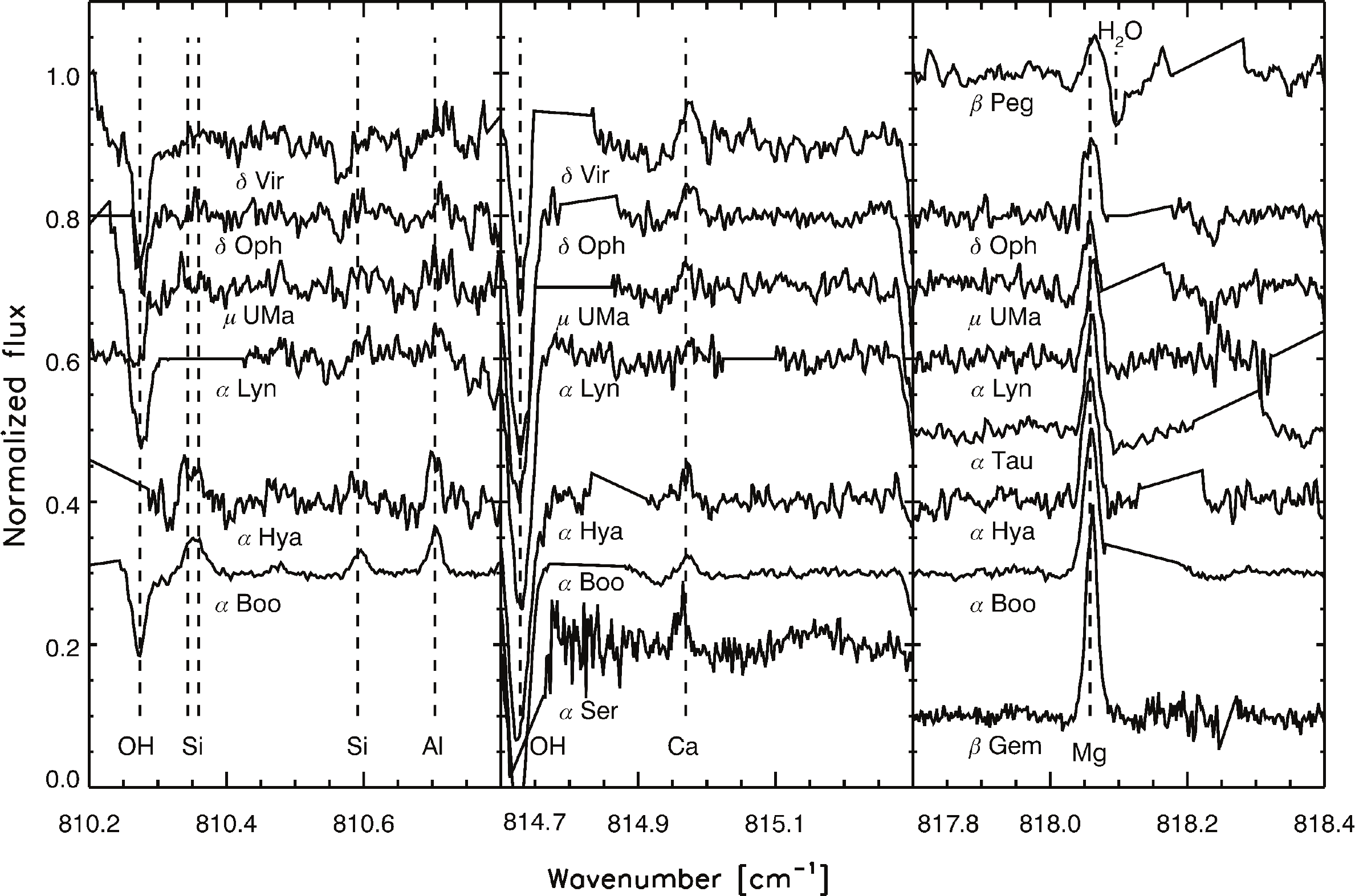}
	\caption{Normalized spectra shifted with respect to each other showing emission lines from Si, Al, Ca, and Mg, which are marked. Two OH lines in absorption are also indicated along with one \water\ line in $\beta$ Peg at 818.1 cm$^{-1}$.} 
	\label{em}
\end{figure*}



\subsection{The HF line}

We detect the rotational HF line at 12.21\,\mic\ (819.11\,\invcm) for all the spectra of our stars, except for $\alpha$ Ser, for which the HF line falls beyond the recorded spectrum. 
This  line can be used to derive the fluorine abundance of the stars. This was done by \citet{joensson:14b}, who synthesised these successfully with photospheric models {\it only}. Earlier studies in the literature of the photospheric fluorine abundances in stars have all used the vibration-rotation HF lines at 2.3\,\mic\ instead. 
\citet{joensson:14b} used both sets of lines to determine the photospheric fluorine abundances for our stars. 
They gave consistent abundances, indicating that also the 12 \,\mic\ HF lines are photospheric in origin and well described by a standard model atmosphere. 
Again, this indicates that we are recording the photospheric spectrum of the red giants at these wavelengths.

\subsection{Velocity shifts between water, OH, and  emission lines}

Table \ref{tab:vel} presents the mean velocity shifts between the water lines, OH($\nu=0-0$)
lines, and the emission line of Mg relative to the HF line. The latter two are most certainly photospheric in origin.
The uncertainties given there are the standard deviation of the mean for the cases where we can measure several lines. We estimate the random measurement uncertainties to $\pm0.5\,$\kms\ and the systematic uncertainties (mainly due to the wavelength calibration) to lie in the range of $\approx 0.5-1\,$\kms. Since the OH lines lie at lower wavenumbers than the Mg and HF lines, these systematic uncertainties might be the reason that the OH-line shifts for all the stars are systematically negative, but within the estimated uncertainties. For a few stars, such as $\alpha$ Tau and $\beta$ Peg, there might be a formal systematic offset between the \water\ lines and the OH lines, but not between the photospheric HF line and the \water\ lines. However, in the case of the $\alpha$ Tau spectrum, the uncertainties are in the range of this offset. The photospheric Mg emission line, for example, is shifted by the same magnitude but in the opposite direction. The same trend is seen in other stars but within the estimated uncertainties.  $\beta$ Peg  is the coolest star, which means that molecules start contributing severely and start blending lines. This leads to larger systematic offsets in the wavelength determinations.
Thus, we conclude that at this accuracy, within $\pm 1.5\,$\kms, all the spectral lines are at the same velocity. 

This means that either the water lines in all our stars are formed in the photospheres or that the MOLspheres have no velocity relative to the stellar photospheres.

\begin{table}
\caption{Velocity offsets, and standard deviations of the mean when there are several lines, in km\,s$^{-1}$ of the water, OH, and  Mg lines relative to the HF line, are shown in columns 2, 3, and 4, respectively. \label{tab:vel}}
\centering
\begin{tabular}{cccc}
\hline  \hline
 Name  & $\Delta v_{water}$ & $\Delta v_\mathrm{OH}$  & $\Delta v_\mathrm{Mg}$  \\
  \hline
 $\beta$ Peg   & $-0.3\pm 0.6$  &$-1.9\pm 0.3$ & 0.3 \\
 $\delta$ Vir  & $0.2 \pm 0.6$  &$-0.6\pm 0.1$ & 0.0  \\
 $\delta$ Oph  & $0.0 \pm 0.3$  &$-0.9\pm 0.5$ & -0.2 \\
 $\mu$ UMa     & $0.5 \pm 0.5$  &$-0.5\pm 0.5$ & 1.3  \\
 $\alpha$ Lyn  & $0.1 \pm 0.5$  &$-0.8\pm 0.2$ & 0.9 \\
 $\alpha$ Tau  & $0.0 \pm 0.3$  &$-0.9\pm 0.1$ & 0.7 \\
 $\alpha$ Hya  & $-1.5^{(a)}$         &$-0.5\pm 1.0$ & 1.2 \\
 $\alpha$ Boo  & $0.0\pm 0.8$   &$-0.7\pm 1.2$ & 1.5 \\ 
 $\beta$ Gem   & $-$            &$-1.7\pm 0.5$ & -0.5\\
 \hline
 \end{tabular}
\tablefoot{\\
\tablefoottext{a}{Only one weak \water\ line}\\
}
\end{table}

\section{Discussion}

Previously, water lines in the mid-infrared have been investigated in only four stars, namely  the red giant $\alpha$ Boo \citep{ryde:water0} and  the red supergiants $\alpha$ Sco \citep{antares}, $\alpha$ Ori \citep{ryde:water2}, and $\mu$ Cep  \citep{ryde:water1}. The question is whether the properties of the strong water lines are specific to a few particular stars or whether they are a general feature of red giants  and supergiants. Our goal is to present empirical evidence for how the strengths of the water lines 
change with spectral type for a range of K and M giants to test for  a trend with temperature. 
Before discussing the general trends and its possible causes, we start off by discussing the published results from the literature. 

In the case of the well-studied K giant Arcturus, \citet{ryde:water0} firmly establish the mid-infrared OH lines to be photospheric by determining identical velocity shifts and line widths of the vibration-rotation lines at 1.6\,\mic, a definite photospheric feature\footnote{The continuous opacity in cool stars shows a minimum at these wavelengths, implying that the continuum is formed at deeper atmospheric layers than the continuum at other optical and infrared wavelengths.}, and the $12\,$\mic\ rotation lines, as well as successfully synthesising lines of different excitation energies, including the strong \water\ lines at 12\,\mic, based on a modified photospheric model only. 
They find that the outer temperatures have to be lowered by of the order of 10\%, that is approximately  300~K compared to the radiative-equilibrium temperatures of 2000-2500~K, in the layers described by $\log \tau_{500}<-4$ (see Figure 7 of \citet{ryde:water0}). This was a so called semi-empirical model atmosphere, constructed in order to fit the water lines. 
In addition,  the metallic emission lines detected in this region, especially the Mg line emission, is shown to be formed by a photospheric line-formation process \citep{sundqvist:08}, establishing also that the continuum most likely originates in the photosphere. 

Although K giants do not have direct evidence for MOLspheres \citep{tsuji:03}, 
\citet{tsuji:09} find indirect evidence  from lines originating from the fundamental band of CO that is interpreted as suggesting a MOLsphere for Arcturus, nevertheless.
However, the \citet{ryde:water0} observations do not register the molecular layers and therefore either contradict the MOLsphere hypothesis\footnote{Assuming that there is no specific selection of only \water\ and CO in the MOLspheres, but that there is simultaneously hydrogen and other molecules, like OH, available. Note, that in the MOLsphere realisations of \citet{ryde:water1}, the continuous opacity is due to H$_\mathrm{ff}$ and  H$^-_\mathrm{ff}$, which is very small. The continuum from the MOLsphere is therefore expected to be very optically thin, i.e. transparent.} or put new strong constraints on the nature and interpretation of the MOLspheres. With a MOLsphere, it would be unlikely that the velocity shifts, line widths, and degree of excitation would match so closely to the photospheric values. 

It can be noted that \citet{cohen:05} find that the continuum of 
$\alpha$ Boo 
is reasonably well modelled by classical model photospheres in the region of $40-125$\,\mic, that is, wavelengths that sample the upper photosphere. But they also show that it is not until wavelengths beyond 1-3 mm that the continuum  is formed in photospheric regions corresponding to where \citet{ryde:water0} need to start cooling their upper photosphere. This is also a location which is well above the depth of the onset of the coexisting chromospheric temperature rise of the Arcturus model of \citet{al:75}.  Chromospheric free-free continuum radiation also contributes at (sub-)millimeter wavelengths. 


Later, \citet{ryde:water2} and \citet{ryde:water1} also found stronger than expected \water\ lines both in the supergiant Betelgeuse and in the prototypical MOLsphere supergiant $\mu$ Cep \citep{tsuji:2000}. Suggested LTE MOLsphere models in the literature for these stars, based on low-resolution spectra and interferometric measurements \citep{tsuji:2000,ohnaka:04a,verhoelst:06}, would all result in strong emission features at 12\,\mic, which are not detected. Also in these cases, lower temperature structures in the outer photospheric layers, inspired by the discussion for Arcturus  \citep{ryde:water0}\label{aboo} but also by that in \citet{antares},  made  it possible  to model the spectra well. The conclusion of \citet{ryde:water2,ryde:water1} is that also in these cases the water lines are most likely formed in the photospheres, albeit in non-classical ones, and that there is no need for MOLspheres to explain this wavelength region. From the line widths it is suggested that the features are formed close to the photosphere, perhaps in connection with the coexisting, inhomogeneous chromosphere. The column densities of \water\ also fit closely what is expected from the spectra \citep{antares}.

A cool temperature structure is able to synthesise all the spectral features of the $12\,$\mic\ region simultaneously because is the relative response of the different spectral features to the outer, cooler layers due to different heights of formation. While the models that explain this wavelength region are semi-empirical model atmospheres, constructed solely to fit the lines, insights may still be gained from such an exercise. It could, for instance, indicate lower excitation temperatures of the levels of water forming the lines, or indicate  extra cooling. 

The ultimate goal, however, is to find a model that can explain all empirical data. This is not yet the case, since a photospheric temperature structure as suggested by the 12\,\mic\ wavelength region predicts strong TiO bands in the optical, which are not seen in Arcturus \citep{ryde:03}. Furthermore, the MOLsphere model of $\mu$ Cep predicts emission from water bands at 6 and 40\,\mic, as  suggested from low-resolution {\it Infrared Space Observatory, ISO} spectra \citep{tsuji:2000}. This scenario predicts the $12\,$\mic\ lines to be in emission as well. However, \citet{ryde:water1} detect the contrary. High-resolution spectra of the 6 and 40\,\mic\ water bands, resolving the individual lines, will be able to reveal the nature of the emission lines, providing line profiles, excitation temperatures, velocity shifts, turbulent velocities, and the existence of absorption lines or components.
Indeed, 6\,\mic\  observations have now been performed during the science verification flight in April 2014 of the EXES spectrometer onboard the {\it Stratospheric Observatory For Infrared Astronomy, SOFIA} \citep[see][]{sofia:exes:SV}. No results have been presented yet.

\subsection{New TEXES observations}\label{discussion}

We begin by summarising the  results based on our new observations at high spectral resolution in the mid-IR; 
the water vapour lines in all our stars from  $\alpha$ Boo and cooler show  absorption lines stronger  than expected from a classical model photosphere. 
The continuum is most likely formed in the photosphere, as is the HF line, metallic emission lines, and  the OH($\nu=2-2$) and OH($\nu=3-3$) lines. What is then the cause of the strong absorption of only the water lines and the 
OH($\nu=0-0$) lines? Note, that these show no net velocity shift compared to the photospheric lines.

\subsection{The general behaviour across spectral types}

The water lines we detect for giants of effective temperatures spanning the range from 3400 to 4900\,K show a smooth trend in the strengths of the water lines. The strong lines and lack of emission, therefore, seems to be a general feature of red giants. An explanation of the stronger-than-expected lines must therefore be found in a general cause, applicable to all late-type giants.  The general nature of the observed molecular mid-IR lines, underpredicted by synthetic spectra, was also found  by \citet{Sloan:14} (in press)  for all K giants in their sample. 
For the cases of the supergiants Betelgeuse and  $\mu$ Cep \citet{tsuji:06,tsuji:09} suggested a specific solution to explain the strong \water\ lines at 12\,\mic\ shown by \citet{ryde:water2}. In this case, optically thin 12\,\mic\ lines were suggested to be formed in the MOLsphere,  in absorption against a continuum formed from an alumina shell. Our stars do not have surrounding dust. 


\subsection{Excitation temperatures of the water lines}

In the wavelength recorded, we actually have water lines with quite a spread in excitation energies of the lower levels of the transitions: 0.398-1.398 eV, see Table \ref{tab:WaterMolData}. In the discussion of $\mu$~Cep \citep{ryde:water1}, the range was smaller: 1.014-1.150 eV. This means that 
we can derive the temperature of the site of line-formation from line-strength ratios. The dotted lines in Figures \ref{data1} and \ref{data2} mark the lines that can be used in the calculation of 
such temperatures.





In order to 
derive analytical relations, we make the following hypothesises: 
\begin{enumerate}
\item For a given star, all water lines are formed in layers with a constant temperature. As the outer parts of static 
red giant atmospheres, where molecular line contribution functions are believed to peak, exhibit a rather flat 
temperature profile, this hypothesis should be rather valid (note however that resonance lines have  relatively broad 
contribution functions). 
\item The continuum opacity is constant over the considered wavelength range. In the infrared, the main continuum 
opacity source is due to H$^-$ free-free absorption, which approximately varies as $\nu^{-2}$, $\nu$ being the frequency. The water lines 
considered here cover a range between 808 cm$^{-1}$ and 821 cm$^{-1}$, i.e. a relative variation of only
$\sim$ 1.5\%. This hypothesis is thus fully justified. 
\item We assume that all lines are formed in the optically thin limit, that is, lines are not saturated; in other words, 
they are located in the linear part of the curve of growth (weak-line approximation). 
\item The lines are not blended. This gets increasingly untrue for the cooler stars. Some lines are blended by as much as 40\%.
\end{enumerate}
The two latter are the most questionable hypotheses.
In spite of the simplifying assumptions, we will calculate the temperatures of the line forming regions. Under these assumptions, the equivalent widths of water lines obey to the relation 
\begin{equation}
\log \left( \frac{W}{\lambda} \right) \, = \, \log C + \log A + \log gf -\frac{5040}{T} \chi 
\end{equation}
where C is a constant parameter (for given star and species), including the continuum opacity and the partition function 
(the constancy of the temperature implies the same partition function and the same molecular equilibrium), $A$ the water 
abundance, and $\chi$ the line excitation energy. 

In other words, line-strength ratios  allows the  determination of  the temperature at which they are formed, or the excitation 
temperature, assumed uniform, in case of non-LTE effects. Considering two lines with excitation energies and oscillator 
strengths $\chi_1$, $gf_1$, and $\chi_2$, $gf_2$, respectively, one gets 
\begin{equation}
T_\mathrm{form} \, = \, \frac{5040 (\chi_1 -\chi_2)}{\log gf_1 - \log gf_2 -log (W_1/W_2)} 
\label{Tform}
\end{equation} 
where $W_1$ and $W_2$ are the measured equivalent widths.

\begin{table}
\caption{Estimated observed ($T_\text{obs}$) and model ($T_\text{synth}$) formation-region temperatures in Kelvin, calculated with the use of Eq.~\ref{Tform}. The difference between the model and observed temperatures, $T_\textit{diff} = T_\mathrm{obs}-T_\mathrm{synth}$, is also computed. Since we assume that all lines are formed in one region and at a single temperature, the mean of all the lines for a given star was obtained. \label{temp}}
\centering
\begin{tabular}{lccc}
\hline  \hline
 Name  &$<T_\textit{obs}>$  &  $<T_\textit{synth}>$   & $<T_\textit{diff}>$   \\
  \hline
 $\beta$ Peg   & 2324  & 2616 & $-277$     \\
 $\delta$ Vir  & 2239  & 2717 & $-478$      \\
 $\delta$ Oph  & 2215  & 3263 & $-1048$             \\
 $\mu$ UMa     & 2369  & 2979 & $-611$              \\
 $\alpha$ Lyn  & 2501  & 3049 & $-537$             \\
 $\alpha$ Tau  & 2257  & 3005 & $-718$             \\
 $\alpha$ Hya  & 2666  & 3262 & $-568$             \\
 $\alpha$ Boo  & 2668  & 3331 & $-663$             \\ 
 \hline

\end{tabular}
\end{table}

Table \ref{temp} displays the obtained values of the temperatures of the line-formation regions both from the observed lines-strength ratios and those from synthetic spectra. The standard deviation due to temperature determination from different combinations of line ratios lie between $100-300$\,K, originating mostly from the simplicity of the assumptions of non-saturated lines and non-blended lines. The temperature from the observed spectra lie around 2500 K and systematically show cooler temperatures compared with the synthetic spectra by 300-1000\,K. Thus, either an additional 
cooling of the outer atmospheric layers, or an additional extension of the atmosphere, compared to the static, LTE  model atmosphere such as a MARCS model, is required to explain the relative strengths  of the
water lines. Figure \ref{aboo_eric} illustrates this for the case of $\alpha$ Boo. 

These temperatures should be seen as indicative. For example, the reason for the large temperature difference for Arcturus (600 K), compared to what was found by \citet{ryde:water0} (300-350 K), comes from the fact that the observed water lines are  weak in the spectrum leading to very uncertain line ratios. The line ratios from the synthetic spectrum are,  however, quite accurate since these lines are not saturated and not blended in this star, being the warmest in our sample. Also, the synthetic spectrum is noise-free.

In any case,  what is found is that the formation temperatures are in general of the order of $300-600$~K cooler as derived from the observed spectra compared to those derived from the modelled ones. This is also consistent with the temperature decreases (300-400 K) needed in the model atmospheres of Arcturus \citep{ryde:water0}, Betelgeuse \citep{ryde:water2}, and  $\mu$ Cep \citep{ryde:water1}.   Also, it is found to be a general feature for all the giants observed here, from K to M giants. It should be noted that the derived temperature could also be the temperature of the MOLsphere if
the water lines are formed in such a circumstellar environment.

\begin{figure}
  \centering
	\includegraphics[angle=-90,width=0.5 \textwidth]{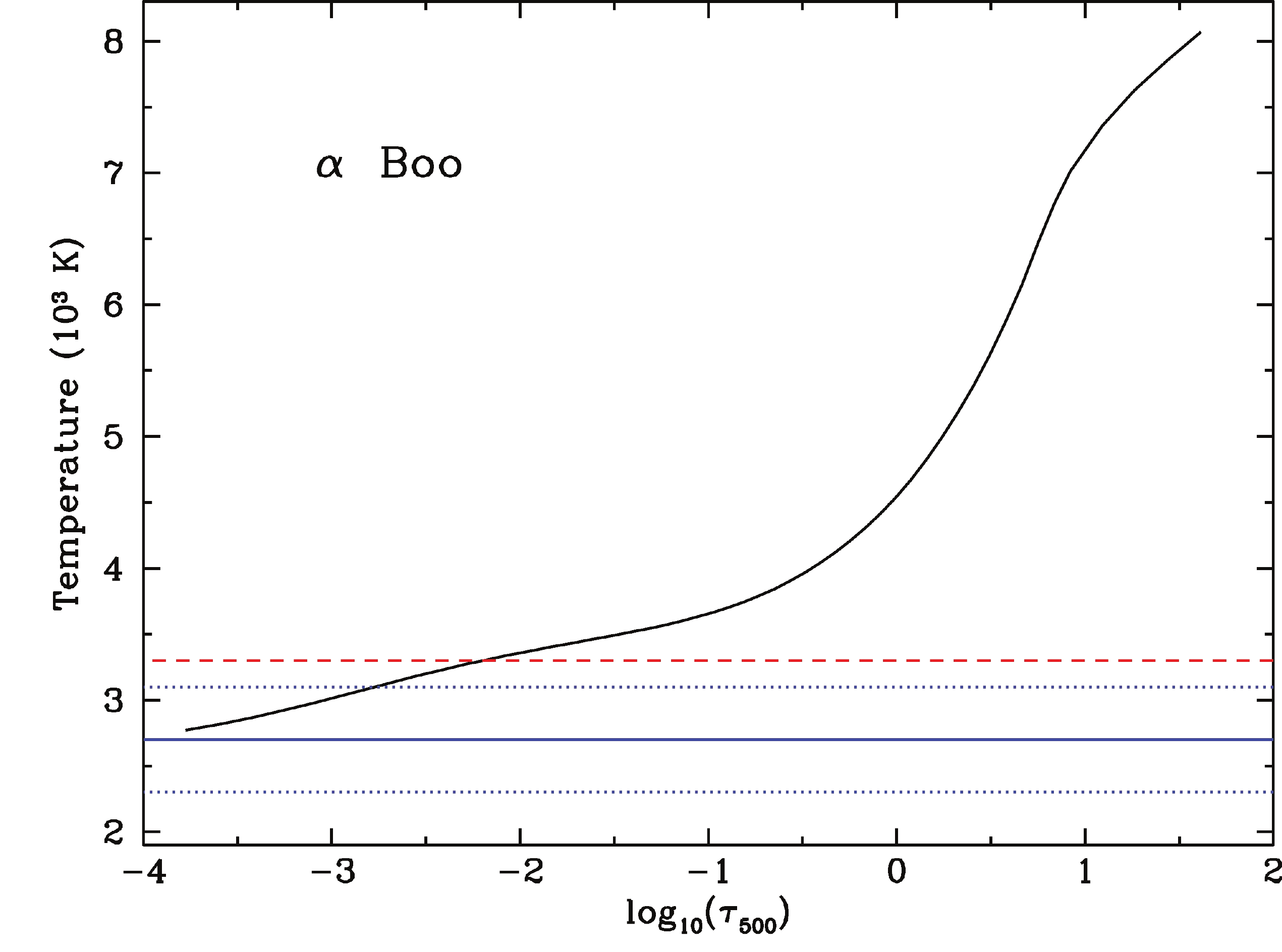}
	\caption{Temperature profile from a MARCS model atmosphere with parameters relevant for $\alpha$ Boo. 
The (red) horizontal dashed line indicates the expected water line formation temperature, assuming LTE.  
The (blue) full line indicates the average formation temperature computed from Eq. \ref{Tform} (the dotted lines 
indicate the full range of temperatures). \label{aboo_eric} } 
	\label{alphaBoo}
\end{figure}

\subsection{Detected photospheric features}

As we have discussed, the observed metallic emission lines at 12\,\mic\ are without doubt photospheric in origin in all our stars. They are formed though a very specific non-LTE process giving rise to them. As mentioned above, they are well explained with this photospheric process for the Sun and giants \citep[see][]{sundqvist:08} anchoring them to the photosphere.
The HF line is  also undoubtably photospheric, since it gives the same abundance as near-IR HF lines \citep{joensson:14b}, as expected from a photosphere. Were they formed in a cooler MOLsphere, the derived abundances would not have matched that closely. 
 The \water\ lines in this wavelength region are, however,  poorly modelled.
As mentioned, Ryde et al. (2002, 2006) thus constructed a semi-empirical model atmosphere which could explain
the formation of strong water lines. A cooling of
the outer atmosphere of a few 100 K, at $\log \tau_{500}<-4$, was needed.  
This extra outer cooling does not, however, significantly affect the HF
line, since it is formed deeper in the photosphere:
the derived  fluorine abundance is only 0.07 dex
lower (see Figure \ref{hfh2o}).

\begin{figure}
  \centering
	\includegraphics[angle=90,width=9.0cm]{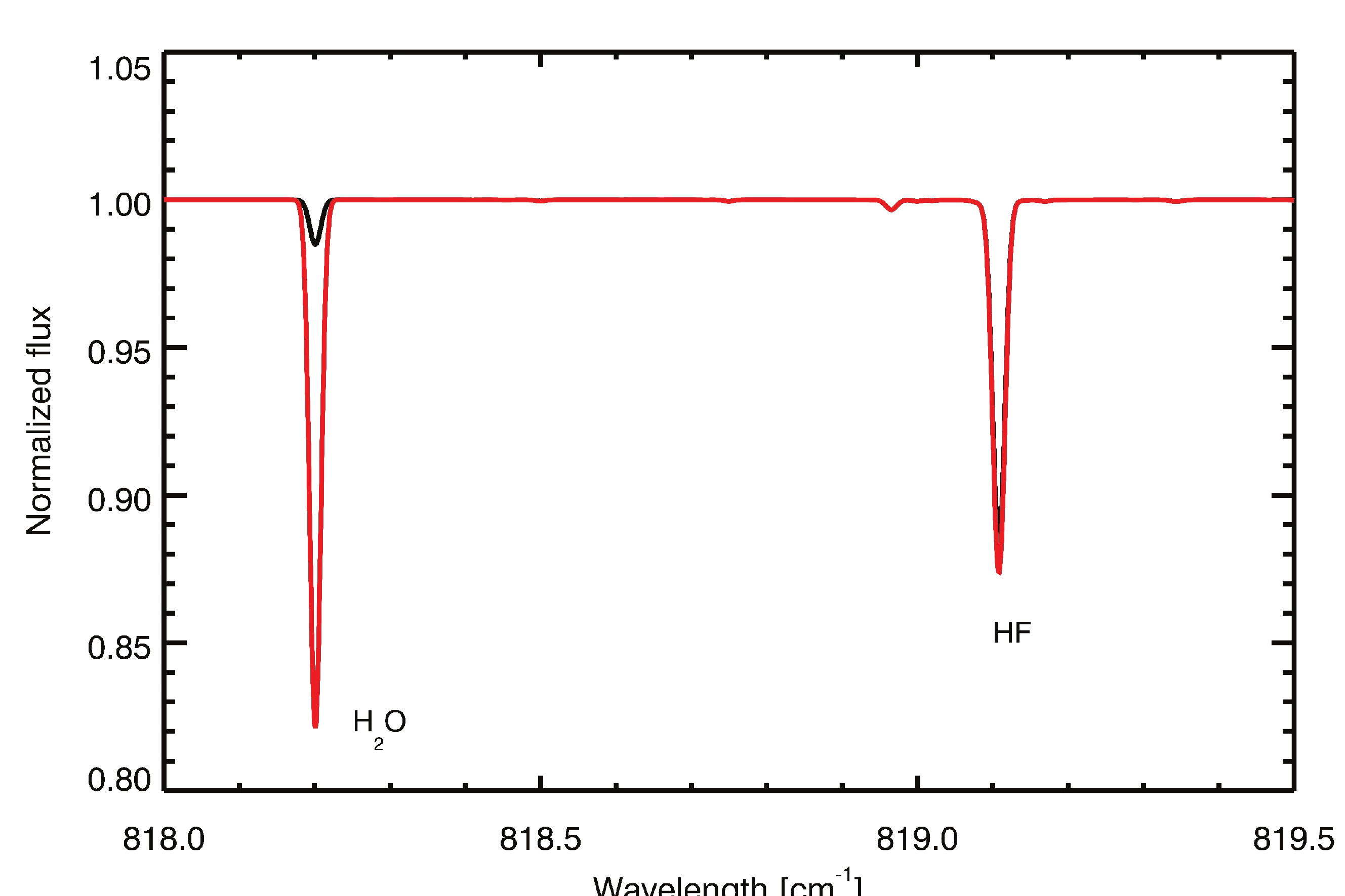}
	\caption{Absorption lines of \water\ and HF calculated based on a non-modified Arcturus model atmosphere (black spectrum) and based on a semiempirical model constructed to fit observed water lines with cooler outer regions (red spectrum), see 
\citet{ryde:water0}. The HF line is formed deeper in the atmosphere and is therefore not affected by the cooler outer parts.} 
	\label{hfh2o}
\end{figure}


The lack of any velocity shift between the photospheric HF and Mg lines is expected. The lack of any shifts of the OH and \water\ lines means either that these lines are photospheric or that they are formed in a MOLsphere that is static, to within a few \kms, relative to the photosphere. \citet{ryde:water0} argued that also the OH rotation lines in Arcturus are formed in the photosphere. This was based on velocity and FWHM similarities between the rotation OH lines and the vibration-rotation OH lines at 1.6\,\mic\ OH lines, the later being photospheric.
We have no reason to believe that this should not be the case for the other stars too, since Arcturus nicely fits into the trend of lines strengths, and velocities. 

Thus, the metallic emission lines, the HF line, and the adjacent continua, are most likely photospheric in origin.  The wavelength shifts of the water lines are, within uncertainties, identical to those of the photospheric emission lines and HF line.   This indicates that also the water lines might be of a photospheric origin. 

 \subsection{Strong \water\ lines} 

We have shown that the approximate temperature at the location of formation of the water lines are a few hundred degrees lower than expected from a model photosphere. 
This is empirically determined from water lines from different excited levels. If the \water\ lines are formed
in such a stellar photosphere, 
lines in LTE would show a larger absorption than expected. We see this for all our giants.

An important point when discussing the formation of the water lines is that they are  optically thick at these temperatures and with the column densities that are assumed in MOLsphere realisations \citep{tsuji:2000,tsuji:03}.  At the line center, $\nu_0$, assuming a  broadened line with a characteristic line width of
$\Delta\nu = \nu_0/c \times (2kT/m_\mathrm{H_2O}+\xi_\mathrm{micro}^2)^{1/2}$, the optical depth is given by 
\begin{equation}
\label{equ}
\tau^l_{\nu_0} \approx 0.02654\times \frac{N_\mathrm{col}}{\sqrt{\pi}\Delta\nu}\, \,(1-e^{-h\nu_0/kT})\,\frac{g_l f_{lu}}{U(1500\,\mathrm{K})}\,e^{-\chi/kT},
\end{equation}

where $N_\mathrm{col}$ is the column density of water vapor, $U$ is the partition function of water vapour and has a value of 2713 for a temperature of 1500 K \citep{par}, and $\chi$ is the excitation energy
of the level \citep[for details see][]{ryde:water1}. For example,  $\log \tau^l_{\nu_0} \approx 2.0$ for
the water line at $818\,$\invcm.

Thus, if the lines were to be formed in a large MOLsphere of these typical temperatures, the sheer size of it will weaken any absorption dramatically, or even show emission \citep[see the discussions in][]{ryde:water2,ryde:water1}. 
Most MOLsphere realisations are of such a size that emission lines are expected at 12\,\mic.  The reason for this is indeed that the water lines
are very optically thick, even in non-LTE, and therefore probe the MOLsphere. The continuum between the \water\ lines, on the other hand, probe the stellar photosphere.  The angular size of a MOLsphere compared to the continuum forming stellar surface, will lead  to absorption lines weaker than expected or even to emission in the lines.    However, we do not find any indication of emission lines of water from any of our red giants, on the contrary they are all deeper than expected.

\subsection{Possible solutions to the mismatch of the \water\ lines} 


Here we discuss possible mechanisms that could increase the strengths of the \water\ lines:

\subsubsection{Spectroscopic uncertainties}\label{solutions}

Increasing the oxygen abundance would increase the strength of the water lines, but it would require an increase of the order of 0.6 dex to fit some of the lines. Other \water\ lines, which are not observed,  would then also emerge. This would also affect the OH lines by a large amount. The oxygen abundances of these stars are quite well determined. For example, \citet{aboo:param} determine the oxygen abundance of Arcturus to better than 0.05 dex. Normally, the abundance determinations are good to within 0.1 dex.

A change in the stellar parameters in the range of their uncertainties will only change the equivalent widths by 20\% for a change in  effective temperature for the coolest star and by up to a factor of 2 for the warmest star, where water is on the verge of being visible. A change in $\log g$ of 0.1 will change the equivalent width of the \water\ lines by 10-20\%. These changes are  too small to explain the strong \water\ lines.

If the lines are saturated, their strengths will be sensitive to the adopted microturbulence. The lines will broaden but they will not appear deeper. A different microturbulence will therefore not be able to fit the strong water lines. 

Thus, none of these uncertainties are enough to explain the lines strengths of the \water\ lines.

\subsubsection{Non-LTE line formation of \water}

The need for considering a situation in which the water lines might not be formed in LTE, was stated already in \citet{ryde:water0}  and discussed by \citet{ohnaka:13:asco}. Different behaviours of rotation lines and vib-rotation lines of \water\ might be caused by different non-LTE effects \citep[see][]{tsuji:06,perrin:04b}. \citet{julien:12} estimated the critical density and showed that it is indeed higher than that of the outer atmosphere or MOLsphere, requiring a proper non-LTE analysis of the formation of water lines either in the photosphere or in a MOLsphere. Preliminary results show that the source function, level populations, and cooling rates are affected (Lambert et al. 2014, submitted, and Lambert et al. 2013b\nocite{julien:12}). 
Either the molecular levels, of for example \water, from which the cooling lines originate (which not necessarily are the lines we have observed) are overpopulated leading to an increased cooling, or the non-LTE level populations could affect the source function and line opacities of the lines discussed. In the latter case the strengths of the lines is a line-formation effect and not an increased abundance of water vapour as such.
%


\subsubsection{Extra cooling in the upper layers of the photosphere}

If we assume that the outer regions in the photosphere are cooler, as indicated by our
line-ratio analysis, how should this additional cooling be interpreted? 
Is it a non-LTE effect or a dynamic effect? 
Non-LTE usually results from either a strong radiation field or a low density of collisional partners. The photospheric radiation 
 field itself may lead to such effects, but it may also be increased by chromospheric emission. This latter would in particular 
manifest itself in the ultraviolet, which 
 would increase the molecular photodissociation, and thus a weakening 
of water lines. We naturally do not rule out the presence of a chromospheric activity in the observed stars, but it does not 
seem to generate a significant  photospheric  non-LTE effect. More generally, a strong radiation field would generate over-excitation (i.e. an 
excitation temperature larger than the kinetic temperature) and over-dissociation, and thus a lowering of the cooling function. 
If non-LTE effects cause the observed strong water lines, it would thus be  mostly through low collision rates compared to radiative ones, leading to a 
stronger cooling. 
An additional cooling  may also be due to a dynamic effect, such as an adiabatic expansion, leading to extra dilution of the outer atmosphere. Radiation pressure, which is potentially sensitive to non-LTE effects, 
 could also produce a similar effect. Furthermore, as the observed giants are believed to be static, non-LTE effects should dominate over dynamic effects.

As mentioned earlier, \citet{julien:13} indeed show that in cool evolved stars, non-LTE effects may be important to the formation of water lines. These effects 
reveal themselves through an overpopulation of lower levels, an increase of the cooling function (the contribution of water to  cooling being multiplied by up to factor of 4 in the outer layers) and increased radiation pressure. This extra water cooling of the outer photosphere could be a reason for the detected cooler temperatures of the line-formation regions. 

\citet{short:03} discussed another reason for a cooler outer photosphere  in their Arcturus model photosphere.  They relaxed the assumption of LTE in the calculation of the photospheric structure by treating atomic opacity (continuous and line opacity) in non-LTE using large model atoms. They demonstrated that overionization of Fe can decrease the temperatures  by a few hundred Kelvin in the outer atmospheric layers. 
It has to be investigated whether this is a general feature for red giants or not. 

In the outer layers where the water lines are suggested to form for all stars across the range of spectral types investigated in this paper, the hot chromospheric regions\footnote{Mg II h/k emission fluxes are reported for all these stars in \citet{perez}}  appear to coexist with cooler areas where water can be formed in sufficient amounts to match our observations. Chromospheres are dynamic and, at least for the solar case, a wave-driven phenomenon of spatially and temporally intermittent structures. The classical modelling neglects mechanical energy and momentum deposition that leads to the chromospheric heating and mass loss, which shows that the classical atmospheric modelling is inadequate for a complete treatment of these regions.  This is in line with the discussion by \citet{ke3} of the COmosphere, that is, the cool regions (determined from CO fundamental lines) coexisting with the chromosphere. They conclude that their CO spectra suggest a dominance of cool regions having large filling factors.

 Temperature inhomogeneities, due to surface granulations in the photospheres,
could affect the water spectrum. In the mid-infrared, the continuum intensity contrast is small. However, the column density of water in the cool, sinking gas would likely be larger than in an homogeneous photosphere. 
The cooler areas in inhomogeneous photospheric models with both cool areas and hotter ones, will mainly affect the mid-IR spectrum. Optical and UV wavelengths are more biased toward hotter regions \citep[see the discussion in ][]{ryde:water2}. Hydrodynamic simulations of granules in giants such as by \citet{Chiavassa:09,Chiavassa:10b,Chiavassa:10a} will help in understanding a more realistic formation of molecular lines in the outer photospheres. The 12\,\mic\ lines provides a good test for three-dimensional model photospheres of red giants and supergiants, which are starting to be realised \citep[see the development in, for example,][]{collet:07,Chiavassa:10b,ludwig:12,Dobro:13}.



\section{Conclusions}

We have observed a range of late-type giants in the N band, at $12.2-12.4\,$\mic, at a high enough spectral resolution so as to resolve individual stellar spectral-lines. 
We detect {\it photospheric} metallic emission-lines (which are formed in non-LTE), OH, and HF lines. In addition we detect stronger-than-expected absorption lines of water vapour  in all the stars showing water lines, that is, for stars cooler than $\sim4300$ K. The strengths of these \water\ lines vary smoothly with the spectral types of the stars. This  shows that an explanation for the formation of these water lines has to generally  apply to all giants across spectral types, ranging 4500 to 3450 K (K1.5-M3.0). 




We have shown that the 12\,\mic\ continuum is formed in the photosphere and that several observed features in the spectra are photospheric in origin (the metallic emission lines, OH lines,  and the HF lines). The water lines in  this continuum, could either be formed in the suggested molecular sphere, a MOLsphere, around the stars, or in the photosphere.
The strong absorption lines of \water, indicating a lack of emission from an extended sphere 
and the lack of any velocity shifts between the photospheric features  and the water vapour lines, indicate that they might be formed close to the stars or in the photospheres rather than in a MOLsphere of a large extent. 

From a simple analysis of the excitation balance of the different water lines, we  show  that, for all the stars across spectral types, they are formed at temperatures several hundred Kelvin cooler than the temperatures expected in the outer photospheres. Thus, if the \water\ lines are formed in the photospheres, the temperature of the outer photosphere of red giants in general need to be cooler.  The physical structure of the outer photospheres of late-type giants ($\log \tau_{500}<-4$) are very uncertain due to low densities and complicated structures \citep[see, for example,][]{tsuji:08,tsuji:09}. 
Small changes in the description  of cooling in the energy equations,  can lead to a change in the temperature structure in tenuous boundary layers where the heat capacity per volume is small. 

Non-LTE treatment of either atomic, molecular, and/or continuous opacities might be necessary. Non-LTE effects in atoms, such as suggested by \citet{short:03}, may provide an explanation to a cooler temperature structure. Other non-LTE effects, which might be important, are those in important molecular opacities affecting the energy balance in the outer boundary layers. Examples are strong water and even CO lines which sample these layers \citep{julien:12}. Non-LTE level populations will directly affect the source function and line opacities, and could also offer a viable explanation for the strong water line absorption.  Inhomogeneties as such are also a source of cooler regions.

The formation of the water lines is still a mystery, but with the high-resolution spectra, we have now gathered empirical facts of the nature of the water lines,  adding new constraints to the emerging picture of the outer atmospheres of red giants 
and that will help solving this issue in the future. 
Note that recently, Sloan et al. (2014) showed that the mid-IR bands of SiO and OH are stronger than expected for K giants too,  based on {\it Spitzer/IRS} spectra. They might thus be probing the same phenomenon described here.




\begin{acknowledgements}
The referee Dr. Greg Sloan is warmly thanked for a very constructive and careful referee's report that improved the paper. N.R. is a Royal Swedish Academy of Sciences Research
Fellow supported by a grant from the Knut and Alice Wallenberg Foundation, 
and acknowledges support from the Swedish Research Council,
VR (project number 621-2008-4245), Funds from Kungl. Fysiografiska S\"allskapet i Lund (Stiftelsen Walter Gyllenbergs fond and M\"arta och Erik Holmbergs donation), and 
financial support from the Swedish Karl Trygger's foundation under grants CTS 12:408 and 13:388 and its investment for science and astrophysics. 
Moreover, we thank the French National Agency for Research (ANR) through program number ANR-06-BLAN-0105, and from "Programme National de Physique Stellaire" (PNPS) of CNRS/INSU, France.  This publication made use of the SIMBAD database, operated at CDS, Strasbourg, France, NASA's Astrophysics Data System, and the VALD database, operated at Uppsala University, the Institute of Astronomy RAS in Moscow, and the University of Vienna.

\end{acknowledgements}

\end{document}